\documentclass{article}

\usepackage{graphicx,epsfig,color}
\usepackage{cite}
\usepackage[english]{babel}
\usepackage{amssymb,amsfonts,amsmath}
\usepackage{verbatim}
\usepackage{tikz}
\usepackage{bbold}
\newcommand{\cA}{{\cal A}}  
  
  \newcommand{\cF}{{\cal F}}

  \newcommand{\cL}{{\cal L}}
  \newcommand{\cN}{{\cal N}}
  
\newcommand{\cQ}{{\cal Q}}

\newcommand{\cW}{{\cal W}}

\newcommand{\be}{\begin{equation}} \newcommand{\ee}{\end{equation}}
\newcommand{\bea}{\begin{eqnarray}} \newcommand{\eea}{\end{eqnarray}}
\newcommand{\beann}{\begin{eqnarray*}}  \newcommand{\eeann}{\end{eqnarray*}}
\newcommand{\bfig}{\begin{figure}} \newcommand{\efig}{\end{figure}}
\newcommand{\ba}{\begin{array}} \newcommand{\ea}{\end{array}}
\newcommand{\bcen}{\begin{center}} \newcommand{\ecen}{\end{center}}
\newcommand{\btab}{\begin{tabular}} \newcommand{\etab}{\end{tabular}}

\newcommand{\Eq}[1]{(\ref{#1})}
\newcommand{\vev}[1]{\left\langle{#1}\right\rangle}

\newtheorem{Proposition}{Proposition}[section]

\newtheorem{Theorem}{Theorem}[section]
\newtheorem{Lemma}{Lemma}[section]
\newtheorem{Corrolary}{Corrolary}[section]

\newcommand{\bp}{\begin{Proposition}}   \newcommand{\ep}{\end{Proposition}}
\newcommand{\bt}{\begin{Theorem}}   \newcommand{\et}{\end{Theorem}}
\newcommand{\bl}{\begin{Lemma}}     \newcommand{\el}{\end{Lemma}}
\newcommand{\bc}{\begin{Corrolary}} \newcommand{\ec}{\end{Corrolary}}

\def\ep{\epsilon}

\def\m{\mu}
\def\n{\nu}

\title{\bf Breaking the sound barrier in AdS/CFT}
\author{Carlos Hoyos${}^{1,a}$, Niko Jokela${}^{2,3, b}$, David Rodr\'{\i}guez Fern\'andez${}^{1,c}$, and Aleksi Vuorinen${}^{2,3,d}$\\
$\ $\\
${}^1${\em Department of Physics, Universidad de Oviedo}\\
{\em Avda.~Calvo Sotelo 18, ES-33007 Oviedo, Spain}\\
${}^2${\em Department of Physics} and ${}^3${\em Helsinki Institute of Physics}\\
{\em P.O.~Box 64, FI-00014 University of Helsinki, Finland}\\
{\small ${}^a$\tt{hoyoscarlos@uniovi.es}, ${}^b$\tt{niko.jokela@helsinki.fi},}\\ 
{\small ${}^c$\tt{rodriguezferdavid@uniovi.es}, ${}^d$\tt{aleksi.vuorinen@helsinki.fi}}
}
\date{}

\begin{document}


{\twocolumn[
\begin{flushright}
FPAUO-16/11 \\
HIP-2016-26/TH\\
\end{flushright}
\maketitle
  \begin{@twocolumnfalse}
  \maketitle
    \begin{abstract}
    \noindent It has been conjectured that the speed of sound in holographic models with UV fixed points has an upper bound set by the value of the quantity in conformal field theory. If true, this would set stringent constraints for the presence of strongly coupled quark matter in the cores of physical neutron stars, as the existence of two-solar-mass stars appears to demand a very stiff Equation of State. In this article, we present a family of counterexamples to the speed of sound conjecture, consisting of strongly coupled theories at finite density. The theories we consider include $\cN=4$ super Yang-Mills at finite R-charge density and non-zero gaugino masses, while the holographic duals are Einstein-Maxwell theories with a minimally coupled scalar in a charged black hole geometry. We show that for a small breaking of conformal invariance, the speed of sound approaches the conformal value from above at large chemical potentials. 
\end{abstract}
  \end{@twocolumnfalse}
]}

\onecolumn

\tableofcontents

\section{Introduction}

Quantitatively understanding the properties of strongly interacting matter in the cold and extremely dense region realized in the cores of neutron stars constitutes a longstanding problem in nuclear physics \cite{Lattimer:2004pg,Brambilla:2014jmp}. The situation is  complicated by a lack of first principles field theory tools: perturbative QCD is only applicable at extremely high densities \cite{Kurkela:2009gj,Kurkela:2016was}, while lattice Monte Carlo simulations are altogether prohibited due to the sign problem \cite{deForcrand:2010ys}. Add to this the fact that robust nuclear physics methods---including their modern formulations, such as the Chiral Effective Theory \cite{Machleidt:2011zz}---are only reliable below the nuclear saturation density $n_s\approx 0.16/\text{fm}^3$ \cite{Tews:2012fj}. It becomes clear that fundamentally new approaches to the problem are urgently needed. In this context, a highly promising avenue is the application of the holographic duality \cite{Maldacena:1997re}, which has indeed been lately applied to the description of both the nuclear \cite{Bergman:2007wp,Rozali:2007rx,Kim:2007vd,Kim:2011da,Kaplunovsky:2012gb,Ghoroku:2013gja,Li:2015uea,Elliot-Ripley:2016uwb} and quark matter \cite{Burikham:2010sw,Kim:2014pva,Hoyos:2016zke} phases inside a neutron star.
 
The most fundamental quantity that governs the thermodynamic behavior of neutron star matter is its equation of state (EoS), i.e.~the functional dependence of its energy density $\varepsilon$ on the pressure $p$. Oftentimes, it is, however more illuminating to inspect the derivative $\partial p/\partial\varepsilon$, which equals the speed of sound squared in the system, $v_s^2$. This quantity namely describes the stiffness of the matter---a property needed to build massive stars capable of resisting gravitational collapse into a black hole. Causality restricts this parameter to obey the relation $v_s^2<1$ (and thermodynamic stability guarantees that $v_s^2>0$), but it has been widely speculated that a more restrictive bound might exist as well. In particular, the lack of known physical systems in a deconfined phase with a speed of sound exceeding the conformal value $v_s^2=1/3$ has prompted a conjecture that this might represent a theoretical upper limit for the quantity  \cite{Hohler:2009tv,Cherman:2009tw} in the same spirit that $\eta/s=1/(4\pi)$  was initially thought to represent a lower limit for the shear viscosity to entropy ratio in any strongly coupled fluid \cite{Kovtun:2004de}. Some support for this argument comes from the fact that both the inclusion of a nonzero mass to a conformal system as well as the introduction of perturbatively weak interactions in an asymptotically free theory are known to lead to a speed of sound below the conformal limit.

The speed of sound conjecture has been widely discussed in the context of neutron star physics, and it has been shown to be in rather strong tension with the known existence of two-solar-mass neutron stars \cite{Demorest:2010bx,Antoniadis:2013pzd}, which requires a very stiff EoS \cite{Bedaque:2014sqa}. This points toward a highly nontrivial behavior of $v_s$ as a function of the baryon chemical potential. Namely, at low densities the speed of sound is known to have a very small value, while its behavior at asymptotically large $\mu_B$ is a logarithmic \textit{rise} toward $v_s^2=1/3$. This implies that should the speed of sound bound be violated somewhere, $v_s$ needs to possess at least two extrema, a maximum and a minimum, between which the quantity may either behave continuously or jump from the maximum to the minimum value.

Recalling the success of holographic methods in the description of strongly coupled quark gluon plasma produced in heavy ion collisions \cite{CasalderreySolana:2011us,Brambilla:2014jmp}, it is clearly worthwhile to study the behavior of the speed of sound in holographic models of quark matter. Here, one, however, quickly realizes that the speed of sound bound is not easily violated: all known examples of asymptotically AdS${}_5$ geometries predict $v_s^2\leq 1/3$  \cite{Cherman:2009tw}.\footnote{For two exceptions to this that are, however, dynamically unstable, see \cite{Buchel:2009ge,Buchel:2010wk}.} The known violations of the bound occur in theories that do not flow to a four-dimensional conformal field theory (CFT) in the UV, and thus do not correspond to ordinary renormalizable field theories in four dimensions. Such examples include the $3+1$-dimensional brane intersections $D4-D6$, $D5-D5$, and  $D4-D8$ (the Sakai-Sugimoto model \cite{Sakai:2004cn}), corresponding to the respective speeds of sound $v_s^2=1/2,1,2/5$  \cite{Jokela:2015aha,Itsios:2016ffv,Kulaxizi:2008jx}. It is well known that even after a compactification to $3+1$ dimensions, it is not possible to disentangle four-dimensional dynamics from the additional degrees of freedom that live on the higher-dimensional color branes, and thus the thermodynamic properties may be very different from a {\em bona fide} four-dimensional theory. Another class of examples that violate the bound are nonrelativistic geometries, such as the Lifshitz ones \cite{Kachru:2008yh,HoyosBadajoz:2010kd,Jokela:2016nsv,Taylor:2015glc}, for which a scaling symmetry fixes $\partial p/\partial \varepsilon=z/3$, with $z$ the dynamical exponent of the dual nonrelativistic theory. For any $z>1$, the EoS of the nonrelativistic theory is stiffer than the conformal one. In this case, the violation of the bound is in some sense trivial, since the dual theory is nonrelativistic.

For the reasons listed above, it is very challenging to build a holographic description for dense strongly interacting quark matter that would allow for the existence of deconfined matter inside even the heaviest neutron stars observed. Indeed, in a recent study of hybrid neutron stars by the present authors \cite{Hoyos:2016zke}, where a holographic EoS was constructed for the quark matter phase, it was discovered that the stars became unstable as soon as even a microscopic amount of quark matter was present in their cores. This was attributed to a very strong first order deconfining phase transition in the model, which was ultimately due to the relatively low stiffness of the conformal quark matter EoS, corresponding to $v_s^2=1/3$. These findings are in line with the analysis of \cite{Kurkela:2014vha}, where it was seen that large speeds of sound were necessary to obtain stars containing nonzero amounts of quark matter in their cores.

One possible resolution to the speed of sound puzzle is clearly that the quantity rises to a value $v_s>1/\sqrt{3}$ in the nuclear matter phase, then discontinuously jumps to a low value at a first order deconfinement phase transition, and finally slowly rises toward the conformal limit in the deconfined phase. In the paper at hand, we propose another viable scenario, involving a violation of the speed of sound bound in the deconfined phase and thereby paving the way to the existence of quark matter in neutron star cores.  We do this by constructing a holographic EoS for dense deconfined matter that not only exhibits a speed of sound above the conformal limit, but in addition involves an asymptotically anti-de Sitter (AdS) spacetime. This provides an explicit counterexample to the common lore that asymptotically AdS spacetimes necessitate $v_s<1/\sqrt{3}$,  \footnote{Note, however, that there is no apparent reason on the field theory side to suspect that the speeds of sound in theories that are UV complete would exhibit a universal upper bound. 
} and suggests that there may exist a large class of realistic holographic models for dense deconfined QCD matter that involve speeds of sound significantly above this bound.

Our paper is organized as follows. In Sec. \ref{sec:speed}, we consider a class of models involving an Einstein-Maxwell action with a minimally coupled charged  bulk scalar, and show that the speed of sound bound is violated in it for different values of the charge and the mass of the scalar. In Sec. \ref{sec:Rcharge}, we consider a top-down string theory setup in this class, dual to ${\mathcal N}=4$ super--Yang-Mills theory at non-zero $R$-charge density --and confirm that the speed of sound is larger than the conformal value before the system becomes unstable toward the formation of a homogeneous condensate. In Sec. \ref{sec:conclusions}, we finally discuss the implications of our findings, while Appendixes \ref{app:nearext}, \ref{app:formulas}, and \ref{app:holoren} are devoted to a closer look at some technical details of our computation.

\section{Speed of sound at finite density}\label{sec:speed}

In order to construct a holographic model for a finite density system, in which the speed of sound bound might be violated, we should clearly incorporate both a nonzero charge density and a breaking of conformal invariance. The simplest such model is Einstein-Maxwell gravity minimally coupled to a scalar field (charged or not), for which the action reads
\begin{equation}\label{eq:bulkaction}
S=\frac{1}{16\pi G_5}\int d^5 x\left[R-L^2F^2- |D\phi|^2-V(|\phi|^2)\right]\ ,
\end{equation}
where $L$ will be fixed to be equal to the AdS radius. The field $\phi$ appearing here is a complex scalar with charge $q$, such that the covariant derivative acting on it reads
\begin{equation}
D_\mu\phi=\partial_\mu\phi-iq A_\mu\phi\ .
\end{equation}
The potential $V$ can on the other hand be expanded to quadratic order as
\begin{equation}\label{eq:potential}
V(|\phi|^2) \simeq -\frac{12}{L^2}+m^2|\phi|^2\ ,  \ m^2L^2=\Delta(\Delta-4)\ .
\end{equation}
Following the usual AdS/CFT dictionary, we take the scalar field to be dual to a relevant operator of dimension $1<\Delta<4$, while the gauge field $A_\mu$ is dual to a $U(1)$ conserved current. The equations of motion following from this action are
\bea
&&D_M\left( \sqrt{-g}g^{MN} D_N \phi\right) - m^2\sqrt{-g} \phi =0 \nonumber\\
&&4 L^2 \partial_M\left(\sqrt{-g} F^{MN}\right) = i q\sqrt{-g}g^{MN}\left[\phi^* \left(D_M\phi\right) -\left(D_M\phi\right)^* \phi \right] \nonumber\\
&&T_{MN}^{(A)} + T_{MN}^{(\phi)} = R_{MN} -\frac{1}{2}R g_{MN} \nonumber\\
&&T_{MN}^{(A)} = 2L^2\left(F_{MA} F_N\!^A -\frac{1}{4}F^2 g_{MN} \right)  \nonumber\\
&&T_{MN}^{\phi}  = \frac{1}{2}\left[\left( D_M \phi\right)^*D_N\phi + \left(D_N\phi\right)^* D_M\phi -\left( |D\phi|^2+V\right)\, g_{MN}\right] \ .\label{EOM}
\eea

If the scalar field is turned off, i.e.~$\phi=0$, a homogeneous and isotropic charged state in the field theory has a gravity dual description given by the AdS Reissner-Nordstr\"om (AdSRN) metric
\begin{equation} \label{eq:holomet}
ds^2=L^2\frac{dr^2}{r^2 f(r)}+\frac{r^2}{L^2}\left[ -f(r)dt^2+d\vec x^2\right] \ , \ f(r)=1+\frac{Q^2}{r^6} -\frac{M}{r^4}\ .
\end{equation}
The gauge potential in the AdSRN solution is
\begin{equation} \label{eq:gauge1}
A_0=\mu\left[1- \left(\frac{r_H}{r} \right)^2\right] \ ,
\end{equation}
where $r_H$ is the position of the black hole horizon and
\begin{equation}
M=r_H^4+\frac{Q^2}{r_H^2}\ , \ \mu=\sqrt{3}\frac{Q}{2L^2 r_H^2}\ .
\end{equation}
The Hawking temperature is then
\begin{equation}
T=\frac{r_H}{2\pi L^2}\left(2-r_H^{-6}Q^2\right)\ .
\end{equation}
At the critical value $Q_c=\sqrt{2}r_H^3$ the solution becomes extremal ($T=0$). The horizon radius can be determined by the temperature and the chemical potential to be
\be \label{eq:rH}
\frac{r_H}{L^2} = \frac{T}{6}\left(3\pi + \sqrt{9\pi^2 + \frac{24\m^2}{T^2}}\right) \ .
\ee

The AdSRN solution is dual to a charged state in a CFT, and therefore the speed of sound is fixed to the conformal value $v_s^2=1/3$. In order to deviate from this value we need to break conformal invariance. 
For this, we will consider a nonvanishing scalar field $\phi\ne 0$. The asymptotic expansion of the scalar close to the AdS boundary is
\begin{equation}\label{eq:rexp}
\phi \simeq \left(\frac{L}{r}\right)^{4-\Delta} L^{4-\Delta}\widetilde{\phi}_{(0,0)}+\left(\frac{L}{r}\right)^{\Delta} L^\Delta \phi_{(0,0)}\ .
\end{equation}
In this expansion, the first term has the interpretation as a deformation of the CFT by the relevant operator with a coupling $J\equiv \widetilde{\phi}_{(0,0)}$, while $\phi_{(0,0)}$ determines the expectation value of the dual operator. An explicit breaking of conformal invariance then corresponds to solutions for which $J\neq 0$. 

In order to simplify the analysis, we will restrict to a small breaking of conformal invariance $J/\mu^{4-\Delta}\ll 1$ and/or $J/T^{4-\Delta}\ll 1$. In this case, the amplitude of the scalar field will be small and can be treated as a probe in the AdSRN geometry. Furthermore, we can approximate the potential by the quadratic and constant term, as in (\ref{eq:potential}), and reduce the problem to solving the linearized equations of motion for the scalar field
\begin{equation}
\frac{1}{\sqrt{-g}} \partial_M \left( \sqrt{-g} g^{MN}(\partial_N \phi -i q A_N\phi )\right)-i q A_M g^{MN}(\partial_N \phi -i q A_N\phi)-m^2\phi=0 \ .
\end{equation}

Taking $\phi=\phi(r)$ and defining $\Delta=2+\nu$, the equation of motion in the AdSRN background reduces to
\begin{equation} \label{eq:scalar1}
\phi''+\left( \frac{5}{r}+\frac{f'}{f}\right)\phi'+\left(\frac{4-\nu^2}{r^2f}+\frac{q^2 L^4}{r^4 f^2 } A_0^2\right)\phi=0 \ .
\end{equation}
For computational purposes, it will be convenient to work with a different radial coordinate $u=r_H^2/r^2$ and introduce a parameter $\cQ$ such that $Q=r_H^3\cQ$. In this coordinate, the equation of motion takes the form
\be \label{eq:phiu}
0 =  \phi''+\frac{\zeta_1(u) }{u}\phi' +\frac{\zeta_2(u)}{u^2}\phi \ ,
\ee
where
\bea
\zeta_1 & = & \frac{u+2}{\cQ ^2 u^2-u-1}+\frac{1}{u-1}+2 \nonumber\\
\zeta_2 & = & \frac{u \left[3 q^2 \cQ^2 (u-1)-4 \left(\nu ^2-4\right) \left(\cQ^2 u-1\right)\right]+4
	\left(\nu ^2-4\right)}{16 (u-1) \left(\cQ^2 u^2-u-1\right)^2}\ .\label{eq:zetacoeffs}
\eea
In these coordinates, the horizon is at $u=1$, while the asymptotic AdS boundary is at $u=0$. The expansion close to the boundary is
\begin{equation}\label{eq:uexp}
\phi \simeq \alpha_- u^{1-\frac{\nu}{2}}+\alpha_+u^{1+\frac{\nu}{2}}\ ,
\end{equation}
where $\alpha_-$ and $\alpha_+$ are the coefficients of the non-normalizable and normalizable modes, respectively. They are related to the coefficients in the expansion \eqref{eq:rexp} as
\begin{equation}
J=\widetilde{\phi}_{(0,0)}=\alpha_- \left(\frac{r_H}{L^2} \right)^{2-\nu}\ , \ \phi_{(0,0)}=\alpha_+ \left(\frac{r_H}{L^2} \right)^{2+\nu}\ .
\end{equation}

\subsection{Near-extremal solutions}

Our first goal will be to determine the behavior of the speed of sound in the limit of large densities $\mu\gg T$. The AdSRN geometry will be very close to the extremal solution, and it will deviate only in the neighborhood of the black hole horizon, where in the extremal case the geometry becomes $AdS_2\times \mathbb{R}^3$ while in the nonextremal case there is an ordinary black hole horizon. 

Let us introduce a small parameter $\epsilon\ll 1$, such that the solution becomes extremal when $\epsilon\to 0$: 
\be
 \cQ=\sqrt{2}(1-\epsilon) \ . 
\ee
In the region away from the horizon, $1-u\gg \epsilon$, we can simply set $\epsilon=0$ to leading order and solve the equations of motion. Deviations from the extremal limit can be computed in a systematic way by means of a perturbative expansion in $\epsilon$. 

In the region close to the horizon $1-u\sim \epsilon$ the naive expansion in $\epsilon$ breaks down, since the geometry deviates from the extremal AdSRN. Instead, one can take a near-horizon expansion by introducing a new radial coordinate $v$ defined as
\begin{equation}
u=1-\frac{4}{3}\epsilon v \, ,
\end{equation}
and expanding the equations to leading order in $\epsilon$. The resulting solution will be a good approximation in the near-horizon region $1\gg 1-u$. We impose the condition that the solution is regular at the horizon, which completely determines it up to an overall constant.

A full solution valid throughout the full geometry can be constructed by matching both kinds of expansions in the overlapping region $1\gg 1-u\gg \epsilon$. We give the full details of the calculation and the matching in Appendix~\ref{app:nearextremal1}. The ratio between the coefficients of the normalizable and non-normalizable modes in the boundary expansion \eqref{eq:uexp} is given in \eqref{eq:cnu1}, reading schematically 
\begin{equation}\label{eq:ratioapam}
\frac{\alpha_+}{\alpha_-}= Z_{\nu,q}\frac{1+\beta_1\left(\frac{4\epsilon}{9}\right)^{\lambda/\sqrt{3} } }{1+\beta_2\left(\frac{4\epsilon}{9}\right)^{\lambda/\sqrt{3} } } \ ,
\end{equation}
where 
\begin{equation}
\lambda=\sqrt{\nu^2-\frac{q^2}{2}-1} \ .
\end{equation}
We see that, depending on whether $\lambda$ is real or imaginary, there can be two qualitatively different behaviors. For imaginary $\lambda$, the ratio will have an oscillatory behavior with a period that depends logarithmically on $\epsilon$. On the other hand, if $\lambda$ is real, the terms depending on $\epsilon$ can be neglected in a first approximation. For the rest of this section, we will take $\lambda$ to be real, which imposes a lower bound on the conformal dimension of the scalar,
\begin{equation}\label{eq:dimbound}
\nu^2 > 1+\frac{q^2}{2} \ \ \Rightarrow \ \ \Delta> 2+\sqrt{1+\frac{q^2}{2}}\geq 3\ . 
\end{equation}
This agrees with the condition that the effective mass of the scalar field is above the Breitenlohner-Freedman bound in the AdS$_2$ region of the extremal black hole \cite{Gubser:2008px,Faulkner:2009wj}:
\begin{equation}
m_{\rm eff}^2 R_{AdS_2}^2 \geq -\frac{1}{4} \ , \ R_{AdS_2}^2=\frac{R^2}{12} \ , \ m_{\rm eff}^2=m^2-\frac{q^2}{2 R^2}\ .  
\end{equation}
Assuming the bound is satisfied, the ratio between the two coefficients is, to leading order in $\epsilon$,
\begin{equation}\label{eq:cnu}
\frac{\alpha_+}{\alpha_-}=-3^{\nu }\frac{ \Gamma (1-\nu ) \Gamma \left(\frac{1+\nu}{2}+\frac{\lambda+q}{2\sqrt{3}} \right) \Gamma \left(\frac{1+\nu}{2}+\frac{\lambda-q}{2\sqrt{3}}  \right)}{\Gamma (1+\nu) \Gamma \left(\frac{1-\nu}{2}+\frac{\lambda+q}{2\sqrt{3}} \right) \Gamma \left(\frac{1-\nu}{2}+\frac{\lambda-q}{2\sqrt{3}}  \right)} \ .
\end{equation}

\subsection{Nonextremal solutions}

If the bound of the speed of sound is violated at large densities, it will approach the conformal value from {\em above}.  At zero density, the results of \cite{Cherman:2009tw,Hohler:2009tv} tell us that at zero density and large temperatures the conformal value is approached from {\em below}. Therefore, one expects an interpolation between the two behaviors as the ratio $\mu/T$ is varied from infinity to zero, and in particular there should be a maximal value of the speed of sound at some intermediate value. In order to study this behavior we need to go beyond the near-extremal limit. Another important reason to do it is that it will allow us to study the stability of the solutions against a condensation of the scalar.

In order to construct the nonextremal solutions, we will resort to numerics to solve the equation \eqref{eq:phiu}. We impose regularity of the scalar field at the horizon and shoot toward the boundary, where we read the values of the coefficients for the normalizable and non-normalizable modes.  The parameters of the numerical calculation that we can vary are $\cQ$ in the equations \eqref{eq:phiu} and the amplitude of the scalar solution, $\alpha_-$, in the boundary expansion \eqref{eq:uexp}. They can be used to determine the value of the chemical potential and the temperature normalized by the coupling of the relevant deformation,
\be \label{eq:reducedmut}
\m_r \equiv \frac{\mu}{J^{\frac{1}{4-\Delta}}}= \frac{\sqrt{3}}{2} \cQ \,\alpha_-^{\frac{1}{\n-2}} \ , \ t_r\equiv\frac{T}{J^{\frac{1}{4-\Delta}}} = \frac{2-\cQ^2}{2\pi} \alpha_-^{\frac{1}{\n-2}}\ .
\ee
The value of the normalizable coefficient $\alpha_+$ is extracted from the numerical solution and can be used to determine the expectation value of the dual operator and the components of the energy-momentum tensor.

The way we proceed is the following. We first expand close to the horizon $u=1$ and impose regularity. The scalar field has a Taylor expansion that we truncate at, say, the fifth order 
\be 
\phi_h(u) = \sum_{n= 0 }^5 \phi^{H}_{(n)}\left(1-u \right)^n \ .
\ee
The coefficients $\phi^H_{(n)}$ for $n>0$ are determined by $\phi^H_{(0)}$ and $\nu$, $q$, and $\cQ$ in \eqref{eq:zetacoeffs}. We fix $\phi^H_{(0)}=1$ and use the truncated expansion to give initial conditions for the numerical calculation
\begin{equation}
\phi_n(1-\epsilon_0)=\phi_h(1-\epsilon_0) \ , \ \phi_n'(1-\epsilon_0)=\phi_h'(1-\epsilon_0) \ ,
\end{equation}
where $\phi_n$ is the numerical solution and we take $\epsilon_0=10^{-5}$ as the cutoff in the radial direction. We shoot toward the boundary using NDSolve in Mathematica 10 and find a numerical solution $\phi_n$ defined in the interval $\epsilon_0\leq u\leq 1-\epsilon_0$. We do this for $\nu=1$, $q=10^{-5}$ to study the speed of sound --and for $\nu=1,1.1,1.3,1.6$ and $q=0.5,1,4,5$ and the critical values $q^*=1.47,1.64,1.97,2.44$ for the stability analysis. In all cases we vary $\cQ$ starting at $\cQ=10^{-5}$ with a step $\Delta\cQ=0.005$ and keeping $\cQ<\sqrt{2}$.

Once we have obtained the numerical solution, we extract the values of the coefficients of the normalizable and non-normalizable modes by evaluating the solution at the boundary cutoff $u=\epsilon_0$. For $\nu> 1$ we define the numerical values as
\bea
(\alpha_-)_n & = & \left. u^{\frac{\nu}{2}-1} \phi_n(u) \right|_{u=\epsilon_0} \\
(\alpha_+)_n & = & \left. \frac{1}{\nu} u^{1-\nu}\partial_u\left[ u^{\frac{\nu}{2}-1} \phi_n(u)\right] \right|_{u=\epsilon_0} \ .
\eea
These values are determined by $\cQ$, $\nu$, and $q$. In the numerical calculation, we have fixed the amplitude of the scalar field. Since the equations of motion are linear, we can generate a full set of values by doing a trivial rescaling,
\begin{equation}
\alpha_-=a (\alpha_-)_n\ , \ \alpha_+=a(\alpha_+)_n \ , 
\end{equation}
for some real number $a$. We determine the value of $a$ by fixing the temperature in units of the relevant coupling \eqref{eq:reducedmut} for each value of $\cQ$.

\subsection{Thermodynamics}

Following the usual AdS/CFT dictionary, the free energy (grand canonical potential) is proportional to the renormalized on-shell action in Euclidean signature $F=T S_{ren}^E$.
Since there is a nonzero chemical potential, it will be convenient to work in Lorentzian signature, such that the renormalized on-shell action reads
\begin{equation}\label{eq:renaction}
S_{ren}=\int dt \, L_{ren}=V_3\int dt\, \cL_{ren}\ .
\end{equation}
In (\ref{eq:renaction}) we have used the fact that the backgrounds we study are homogeneous, and $V_3$ is the corresponding spatial volume along the boundary directions.
The free energy density becomes then
\begin{equation}
\cF=-\cL_{ren}\, ,
\end{equation}
which we have computed in \eqref{eq:freeenergy} in terms of the coefficients of the asymptotic expansions given in \eqref{eq:nbseries1}. Comparing with the pressure $p=\vev{T^{ii}}$, given in \eqref{eq:tii}, we see that the free energy is equal to minus the pressure $\cF=-p$. The expressions for the energy density $\varepsilon=\vev{T^{00}}$ and the charge density $n=\vev{J^0}$ can be found in \eqref{eq:t00} and \eqref{eq:j0}, respectively. Using these expressions as well as \eqref{eq:solsbge} and \eqref{eq:betaH}, one finds the usual thermodynamic relation 
\begin{equation}
\varepsilon+p=\mu n+Ts \ .
\end{equation}

It will be convenient to use dimensionless quantities given in units of the scale introduced by the coupling for the scalar operator $J$. We will also omit a common factor that  depends on the AdS radius $L$.\footnote{$L^3/G_5\propto N_c^2$ following the usual AdS/CFT dictionary, so this just amounts to omitting a constant factor proportional to the number of degrees of freedom of the theory.} We then define the reduced quantities:
\be \label{eq:reduced}
\varepsilon_r = \frac{\varepsilon}{\frac{L^3}{16\pi G_5}J^{\frac{4}{2-\nu}}} \ , \ p_r = \frac{p}{\frac{L^3}{16\pi G_5}J^{\frac{4}{2-\nu}}} \ , \ v_r = \frac{\vev{\mathcal{O}}}{\frac{L^3}{16\pi G_5}J^{\frac{2+\nu}{2-\nu}}} \ , \ n_r = \frac{n}{\frac{L^3}{16\pi G_5}J^{\frac{3}{2-\nu}}}\ .
\ee
The energy density and pressure depend on the enthalpy $w=\varepsilon+p$ and the expectation value of the scalar operator as
\bea
\varepsilon_r & = & \frac{3}{4}w_r  +\frac{1}{2} (2-\nu) v_r \\
p_r & = & \frac{1}{4}w_r -\frac{1}{2} (2-\nu) v_r  \\
v_r & = & -2\nu\frac{\alpha_+}{\alpha_-} \alpha_-^{\frac{2\nu}{\nu-2}} \ .
\eea
Note that in the linearized approximation the ratio $\alpha_+/\alpha_-$ is independent of $\alpha_-$.

Comparing the Reissner-Nordstr\"om solution for the background metric \eqref{eq:holomet} with the expansions \eqref{eq:nbseries1} and renormalized values \eqref{eq:t00}, \eqref{eq:tii}, one can write the enthalpy as 
\begin{equation}
w_r=\frac{4M}{J^{\frac{4}{2-\nu}} L^8}=\frac{4r_H^4}{J^{\frac{4}{2-\nu}}  L^8} (1+  \cQ^2)+O\left(\frac{J^2}{J^{\frac{4}{2-\nu}} }\right) \ . 
\end{equation}
Neglecting the subleading terms and using \eqref{eq:reducedmut}, this can be written as
\begin{equation}
w_r\simeq  4\alpha_-^{\frac{4}{\nu-2}}(1+\cQ^2) \ . 
\end{equation}

\subsection{Speed of sound}

There are several possible definitions of the speed of sound in a charged system, depending on which thermodynamic quantities are held fixed. The speed of the sound waves is usually associated with a quantity called the adiabatic speed of sound, but for us it will be easier to compute the isothermal speed of sound. The difference between the two is proportional to the ratio $T/\mu$, so the distinction is unimportant for large values of the chemical potential. In our calculation we can vary the relative values of the temperature and chemical potential with respect to the scale fixed by the symmetry breaking coupling $J$. Through \eqref{eq:reducedmut}, those can be parametrized by variations of $\alpha_-$ and $\cQ$. An isothermal variation $t_r=$ constant will satisfy 
\begin{equation}
\alpha_-(\cQ)=(\pi t_r)^{\nu-2} \left(1-\frac{\cQ^2}{2}\right)^{2-\nu} \ .
\end{equation}

The changes in the enthalpy and the vacuum expectation value (VEV) for large values of the chemical potential are
\bea
dw_r & \simeq  & 24\alpha_-^{\frac{4}{\nu-2}} \cQ \frac{2+\cQ^2 }{2-\cQ^2}d\cQ \\ 
dv_r & \simeq & -2\nu  \alpha_-^{\frac{2\nu}{\nu-2}} \left[ 4\nu\frac{\cQ}{2-\cQ^2}\frac{\alpha_+}{\alpha_-}+\frac{d}{d\cQ}\left(\frac{\alpha_+}{\alpha_-}\right) \right]d\cQ \, ,
\eea
leading to the ratio
\begin{equation}
\frac{d v_r}{d w_r} =  -\frac{\nu}{3(2+\cQ^2)}  \alpha_-^2 \left[ \nu\frac{\alpha_+}{\alpha_-}+\frac{2-\cQ^2}{4\cQ}\frac{d}{d\cQ}\left(\frac{\alpha_+}{\alpha_-}\right) \right].
\end{equation}
The isothermal speed of sound becomes in turn
\begin{equation}
v_s^2=\left( \frac{\partial p_r}{\partial \epsilon_r}\right)_{t_r} =\frac{1}{3}\frac{1-2(2-\nu)\frac{d v_r}{d w_r}}{1+\frac{2}{3}(2-\nu)\frac{d v_r}{d w_r} }\simeq \frac{1}{3}\left(1-\frac{8}{3} (2-\nu) \frac{d v_r}{d w_r}\right).
\end{equation}
At large values of the chemical potential, the solution is near extremal, $\cQ\simeq \sqrt{2}$, and 
\begin{equation}
\alpha_-^2\simeq \left(\sqrt{\frac{2}{3}} \mu_r\right)^{2\nu-4} \ll 1 \, ,
\end{equation}
while the speed of sound becomes
\begin{equation}
v_s^2\simeq \frac{1}{3}\left(1+4 C_\nu  \mu_r^{2\nu-4}\right) \ .
\end{equation}
The coefficient that appears in the correction to the conformal value is
\begin{equation}
C_\nu =\frac{1}{8}\left( \frac{2}{3}\right)^{\nu}\nu^2 (2-\nu)  \frac{\alpha_+}{\alpha_-},
\end{equation}
where the value of the ratio $\alpha_+/\alpha_-$ is given in \eqref{eq:cnu}. The sign of $C_\nu$ determines whether the speed of sound is above the conformal value (positive $C_\nu$) or below (negative). As one can see in Fig.~\ref{fig1}, there are values of $\nu$ and $q$ for which $C_\nu >0$ is possible. In particular, for $q=0$ the speed of sound is above the conformal value for any $\nu$.

\begin{figure}[th!]
\begin{center}
\includegraphics[width=0.75\textwidth]{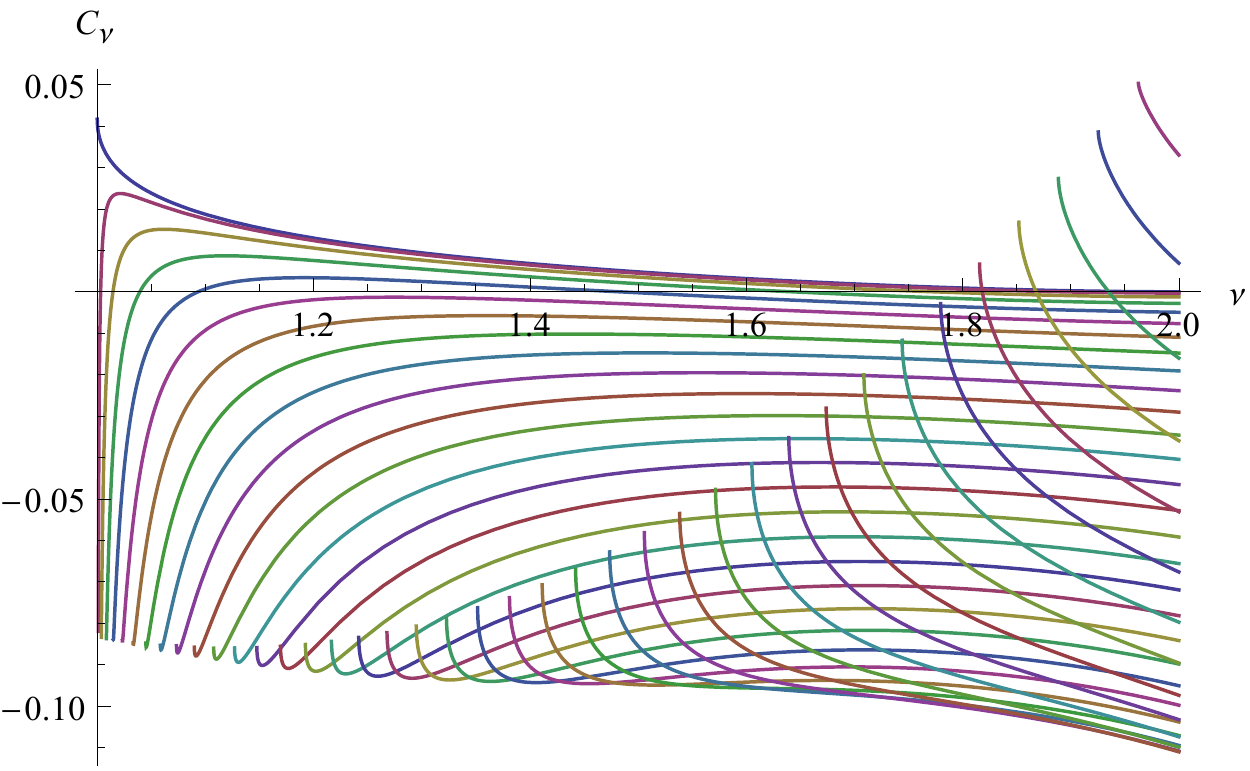}
\end{center}
\caption{$C_\nu$ vs $\nu$  ($1\leq \nu\leq 2$) for different values of $0\leq q < \sqrt{6}$. For $q=0$ (the upper curve), $C_\nu>0$ $\forall \nu$, but as $q$ increases, a range of values of $\nu$ exists  for which $C_\nu<0$. If $q$ is close to $\sqrt{6}$, then $C_\nu$ becomes positive again.}\label{fig1}
\end{figure}

\begin{figure}[t]
	\begin{center}
\includegraphics[width=0.48\textwidth]{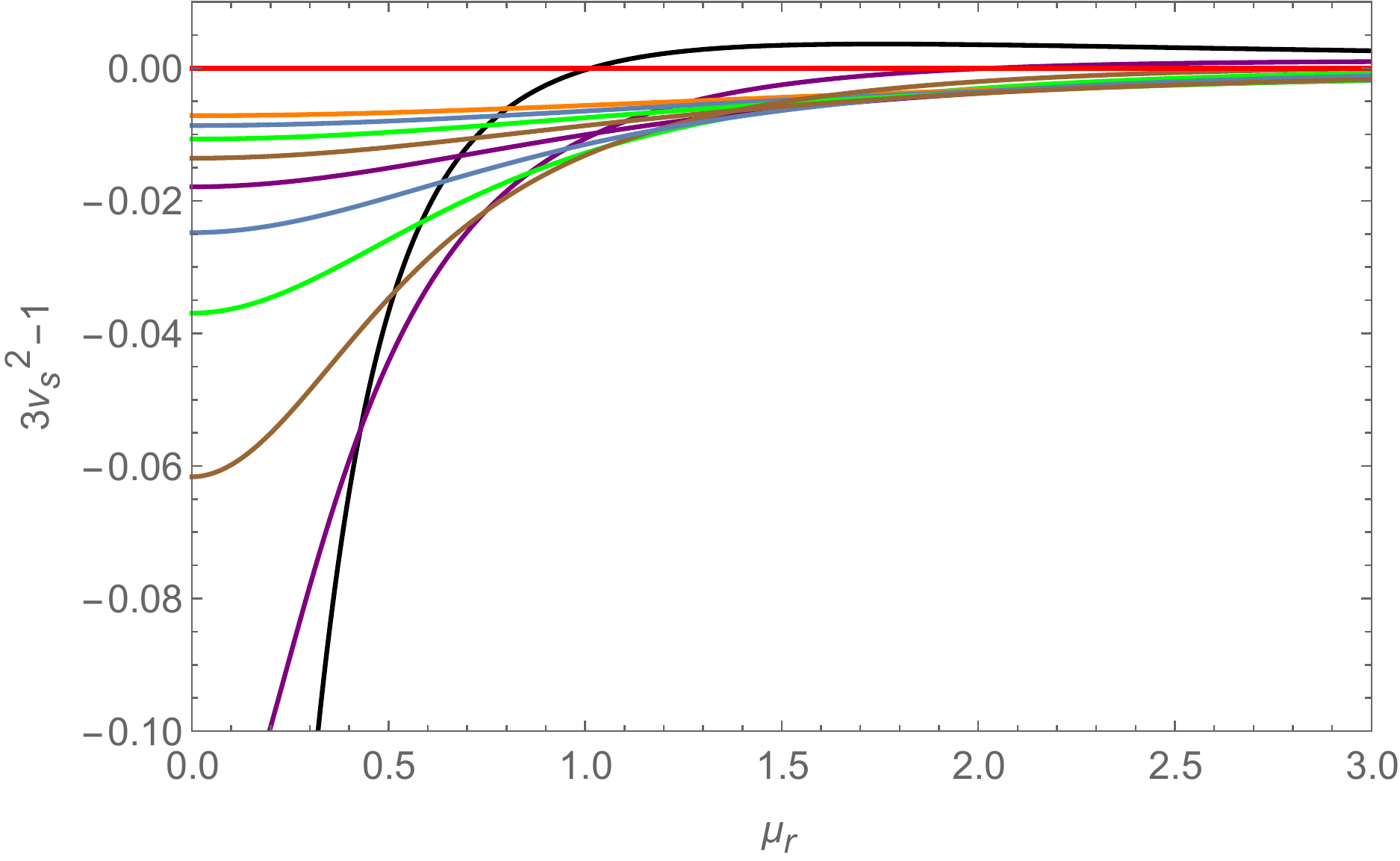}
\includegraphics[width=0.48\textwidth]{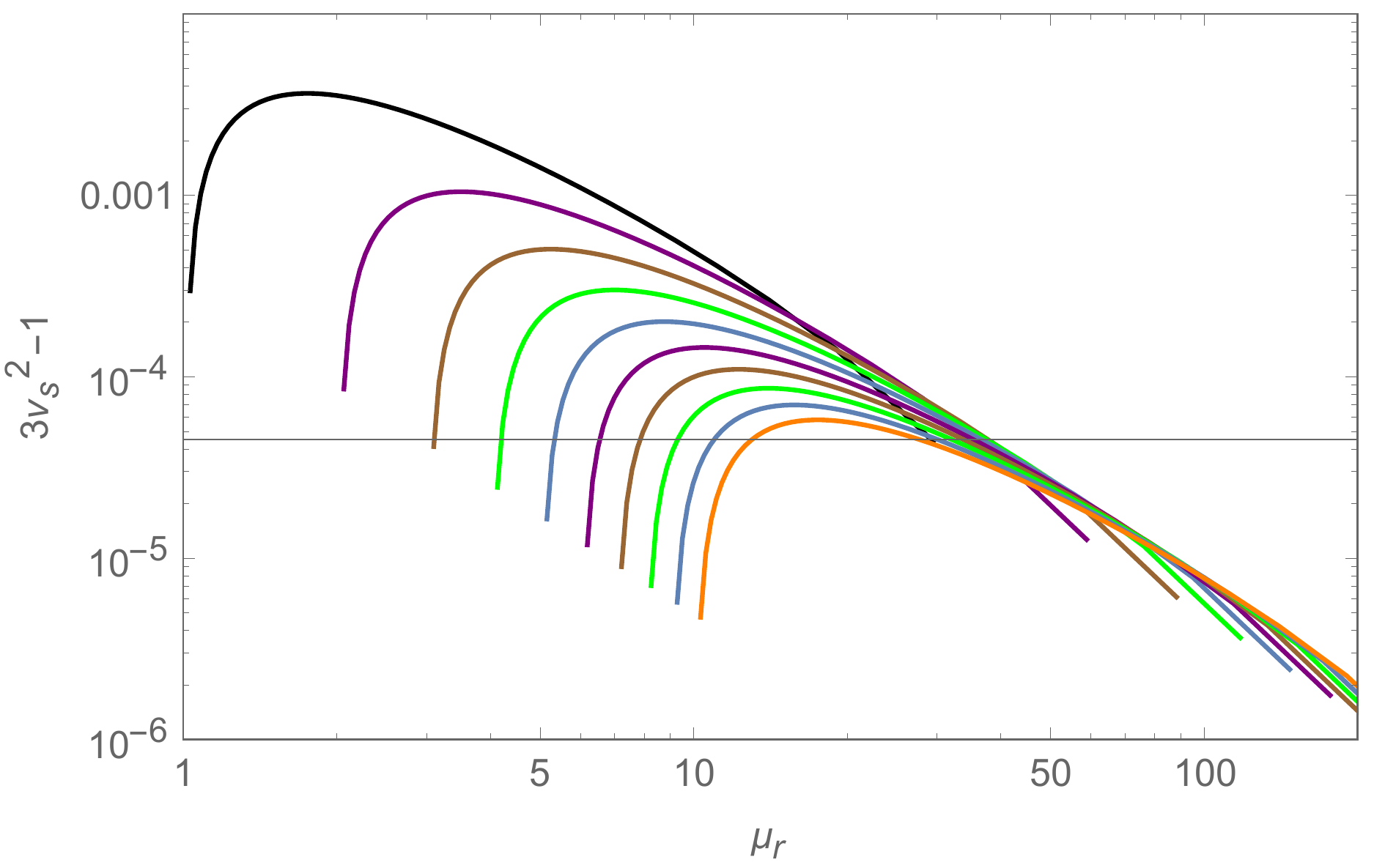}
	\end{center}
	\caption{Left figure: $3v_s^2-1$ as a function of chemical potential for fixed values of the temperature. The black curve (lowest at $\mu_r=0$) is for $t_r=0.1$, and the orange curve (highest at $\mu_r=0$) is for $t_r=1$. Curves in between have intermediate values separated by $\Delta t_r=0.1$ steps. We have taken $q\approx 10^{-5}$,  $\n = 1.1$. Right figure: the same plot in logarithmic scale, from where one can see the maximum more clearly.}\label{fig:csisothermal}
\end{figure}

We have determined the asymptotic form of the speed of sound at very large densities. When $\mu_r/t_r$ is finite, there will be temperature-dependent corrections and eventually, for $\mu_r/t_r\ll 1$ one should recover the results of \cite{Cherman:2009tw,Hohler:2009tv} and find that the speed of sound is always below the conformal value. Using our numerical solutions we plot the speed of sound as a function of the chemical potential for fixed values of the temperature in  Fig.~\ref{fig:csisothermal}. We have chosen $q$ and $\Delta$ in such a way that the near-extremal analysis predicts an asymptotic speed of sound above the conformal value, which can be appreciated in the large-$\mu_r$ part of the plot. For small values of $\mu_r$ we observe that indeed the speed of sound falls below the conformal value. One also notices that as $t_r$ is increased, the speed of sound becomes globally closer to the conformal value, as expected. The position of the maximum in the speed of sound depends strongly on the temperature, and, not surprisingly, it gets pushed to larger values of the chemical potential as the temperature is increased.  The maximum amplitude seems to become steeper as the temperature is lowered, which suggests that the stiffest equations of state would be realized at very low temperatures and chemical potentials of order of the scale set by the relevant coupling.

\subsection{Stability}

We have confirmed the existence of finite density states for which the bound on the speed of sound is violated in holographic models with a UV fixed point. Note that violations were also observed in \cite{Buchel:2009ge} at zero density, but the states for which this happened are unstable \cite{Buchel:2010wk}, so they may be discarded as unphysical. In principle, the same could happen for the states we are considering here, so it is important to check the stability of our solutions. A full-fledged analysis would require a study of the full spectrum of quasinormal modes at finite frequency and momentum. This is beyond the scope of the present paper, but we will analyze stability against the formation of a homogeneous condensate, which is usually the first kind of instability one encounters as the density is increased. This implies that we will restrict the discussion to zero momentum modes. Other instabilities may occur for nonzero momentum that would involve the breaking of translational and/or rotational symmetries; see e.g. \cite{Domokos:2007kt,Nakamura:2009tf,Bergman:2011rf}.

At zero chemical potential and finite temperature the state will be dual to a thermal state of a CFT. If the relevant coupling is zero, $J=0$, the expectation value of the scalar operator will vanish. Technically, there is no solution to the equations of motion for the scalar such that one has a solution which is both normalizable and regular at the black hole horizon at zero frequency (we should remark that we are discussing configurations for which the scalar is close to the critical point of the potential, $\phi=0$). There are normalizable and regular solutions that correspond to the quasinormal modes of the scalar field for complex values of the frequency. Stability of the large temperature state implies that those modes are located on the lower complex frequency plane.

If we turn on the chemical potential and start increasing its value, eventually there could be a point where a zero frequency normalizable and regular solution exists for the scalar. This can be seen as having a quasinormal mode that moves on the complex frequency plane as the chemical potential is varied and reaches the origin. Further increasing the chemical potential typically makes the quasinormal mode migrate to the upper complex frequency plane, thus becoming an instability. Therefore, one can determine the onset of the instability as the point where a regular and normalizable solution appears for the first time. From the field theory perspective this is the critical point that marks the onset of spontaneous symmetry breaking and the formation of a condensate.

\begin{figure}[th!]
\begin{center}
\begin{tabular}{cc}
\includegraphics[width=0.47\textwidth]{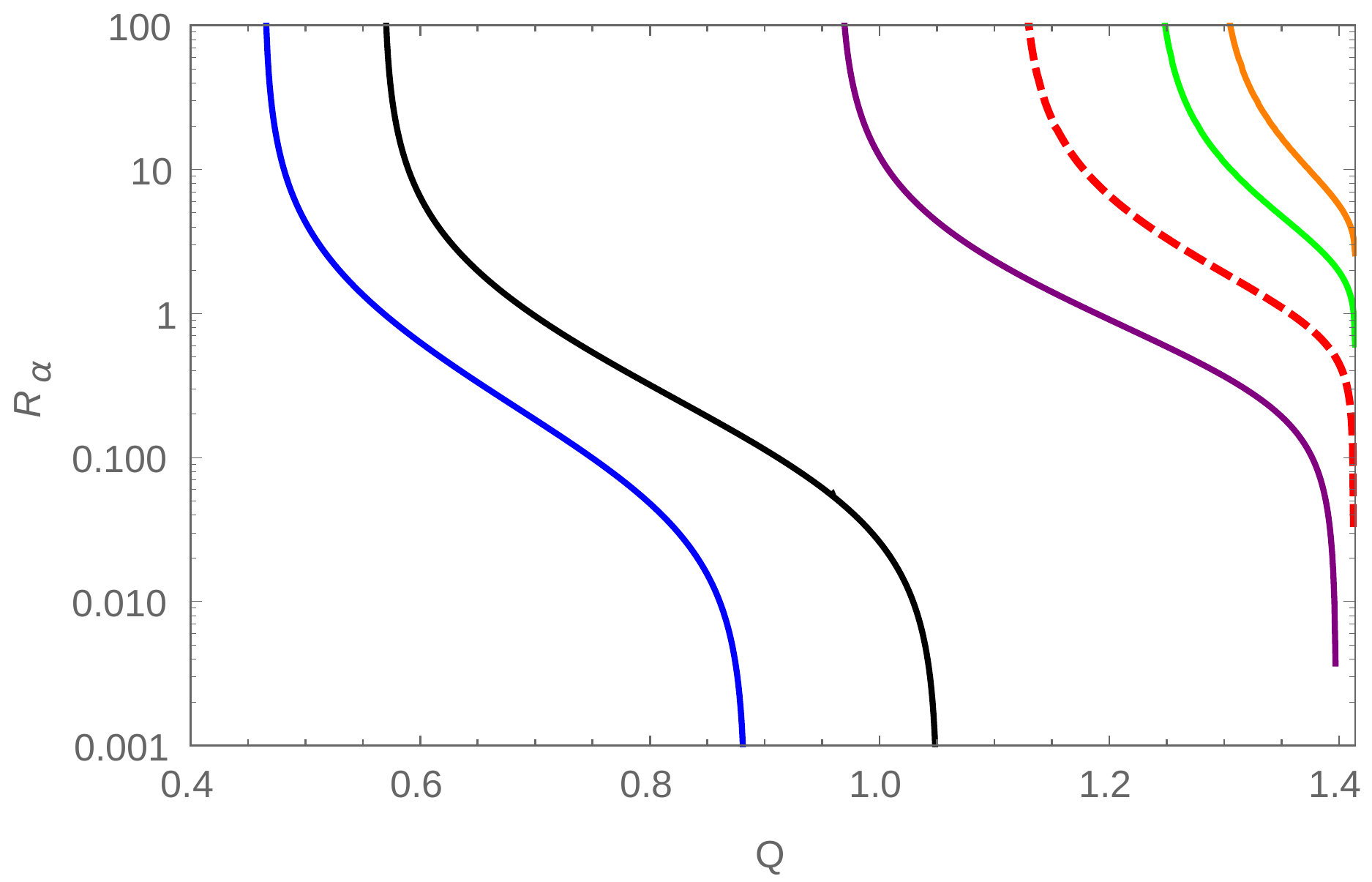} &
\includegraphics[width=0.47\textwidth]{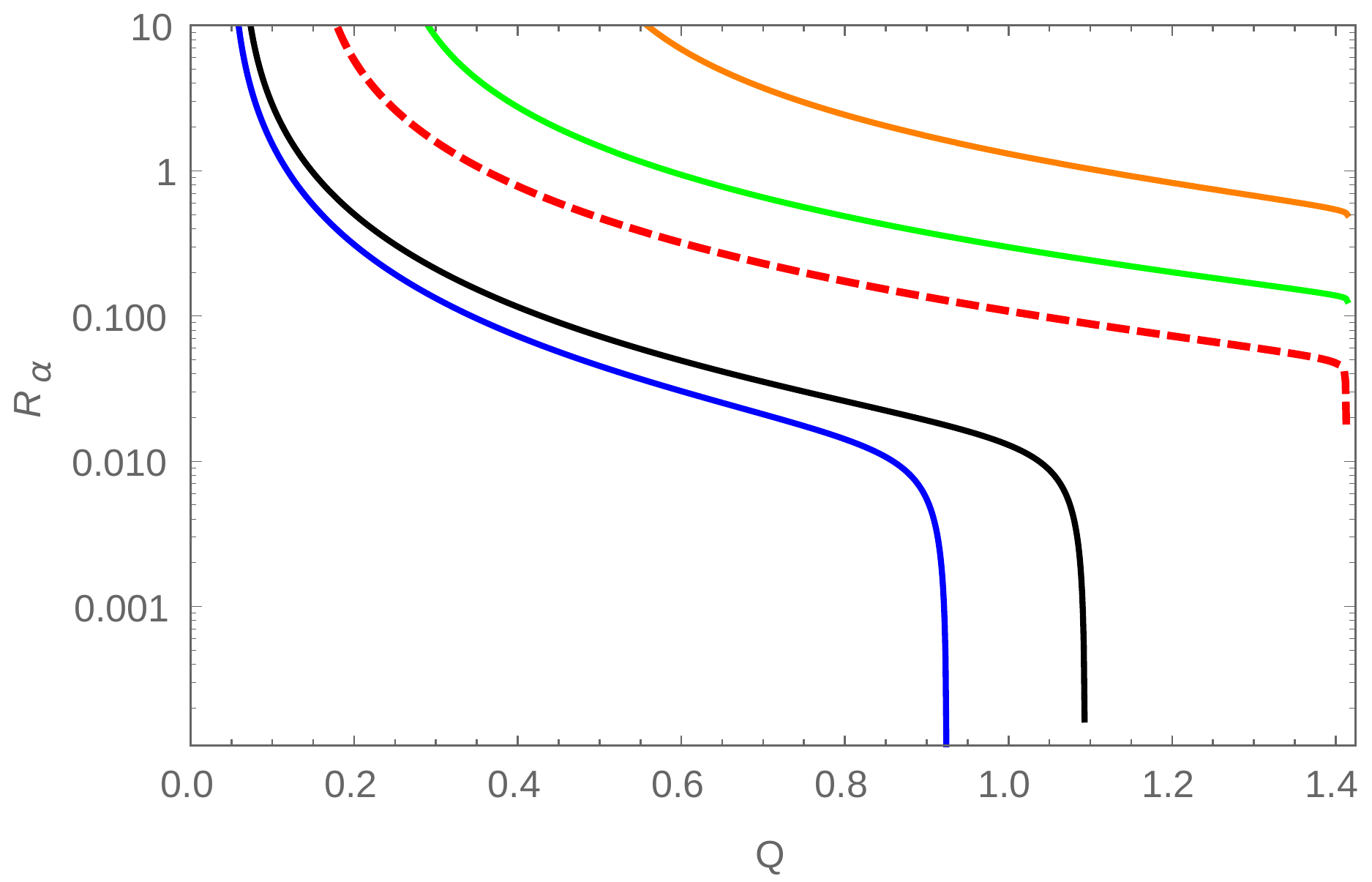}\\
\includegraphics[width=0.47\textwidth]{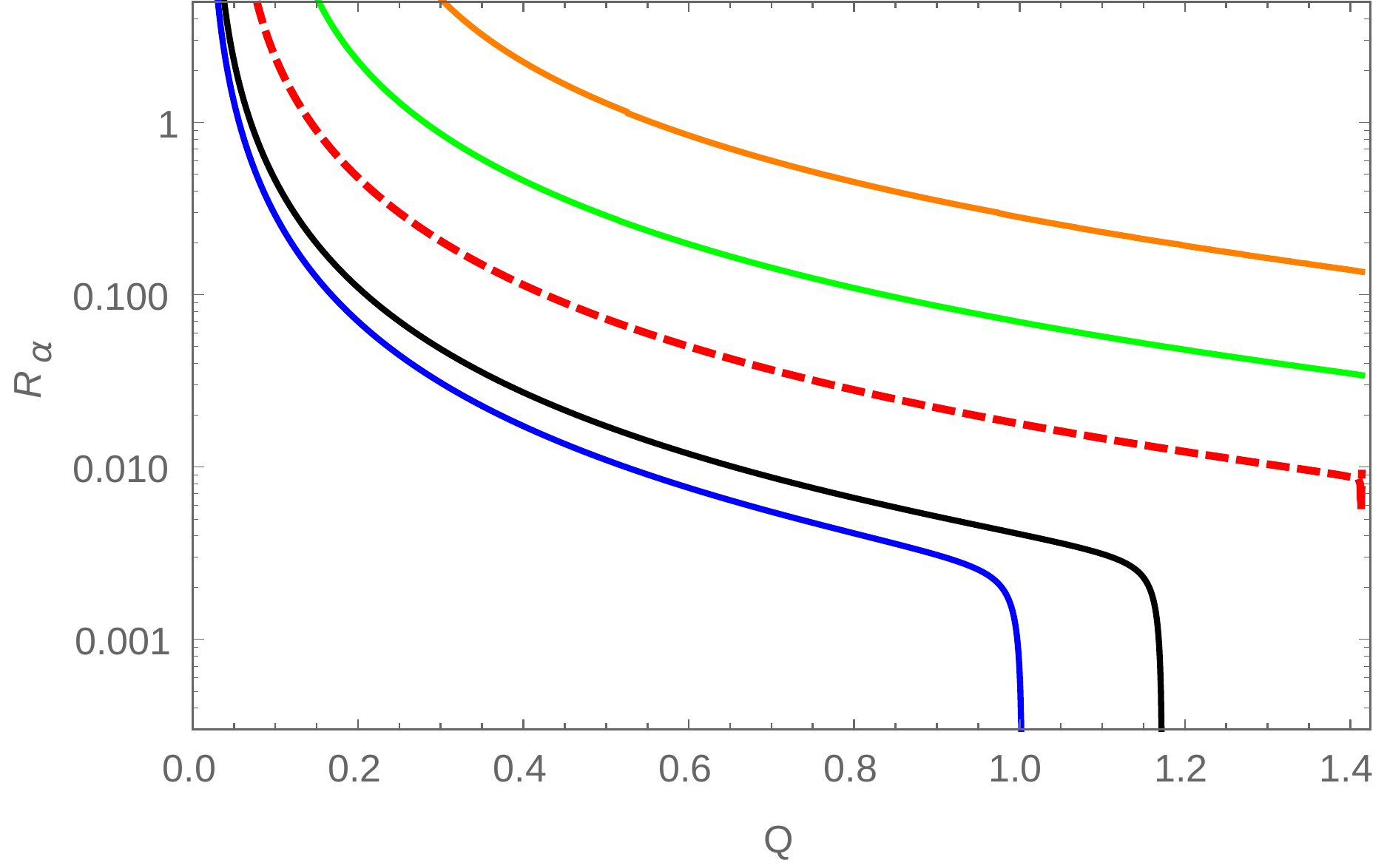} &
\includegraphics[width=0.47\textwidth]{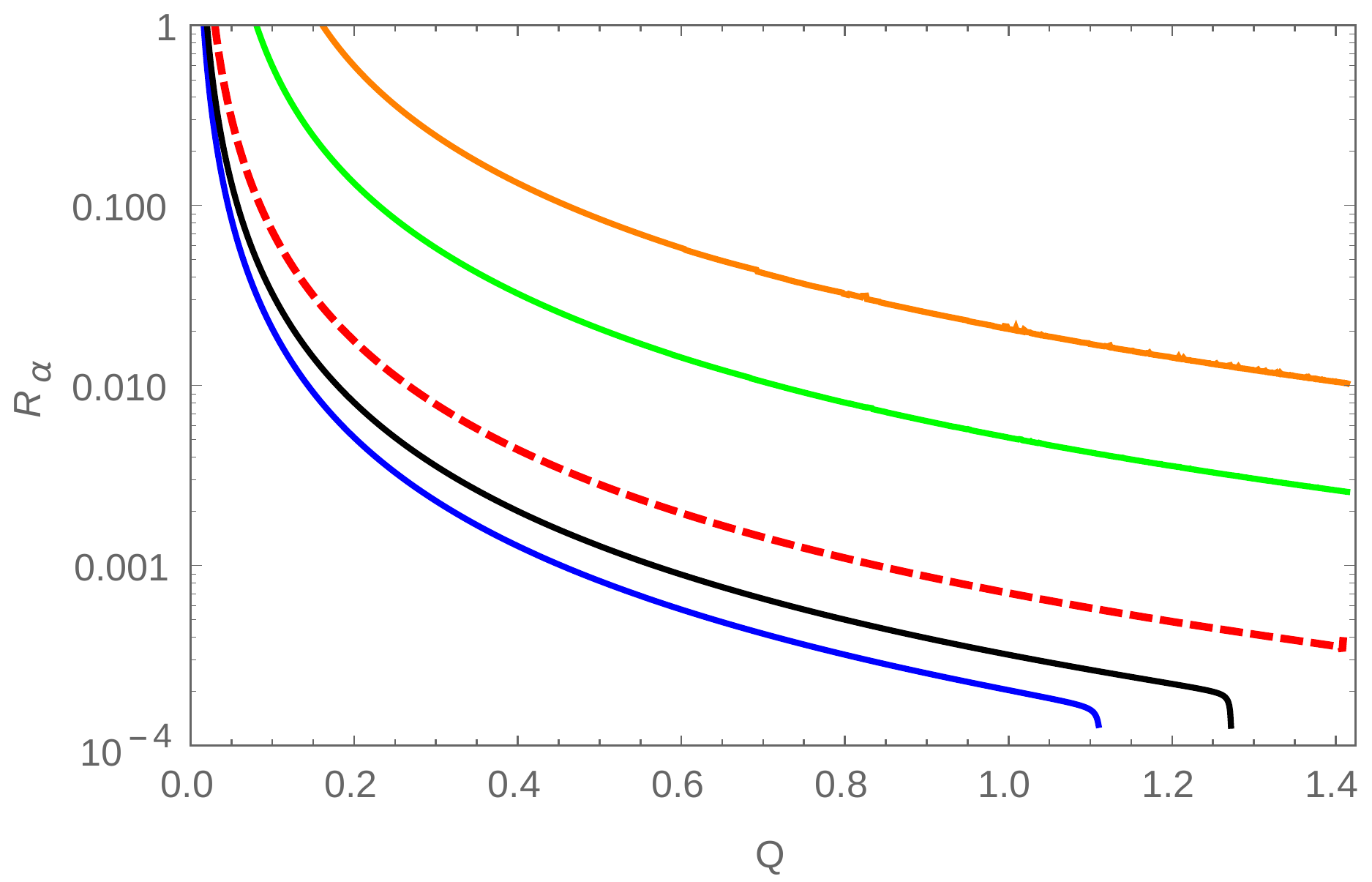}
\end{tabular}
\caption{Ratio of non-normalizable over normalizable coefficients of zero frequency regular solutions $R_\alpha=\alpha_-/\alpha_+$ as a function of $\cQ \propto \m_r/t_r$. The curves represent different values of the charge of the scalar: $q=0.5$ (orange),  $q=1$ (green), $q=4$ (black), and $q=5$ (blue). For $\nu=1$ we have plotted in addition the curve for $q=2$ (purple), corresponding to the string theory model introduced in Sec. \ref{sec:Rcharge}. The red dashed curve denotes the approximate value of the charge $q^*$ at which the first instability appears at $\cQ=\sqrt{2}-10^{-3}$. Each plot corresponds to a different value of the conformal dimensions of the scalar. From left to right and top to bottom:  $\n=1$, $q^*\simeq 1.47$;  $\n=1.1$, $q^*\simeq 1.64$; $\n=1.3$, $q^*\simeq 1.97$; and $\n=1.6$, $q^*\simeq 2.44$. As the conformal dimension is increased, the onset of the instability for a fixed $q$ happens at larger values of $\mu_r/t_r$. \label{fig:ins} }
\end{center}
\end{figure}

The form of the near-extremal solutions suggests that if the bound \eqref{eq:dimbound} is not satisfied, the oscillatory behavior of the coefficients will probably lead to the appearance of zero frequency quasinormal modes and hence instabilities, so this puts a bound on the charge of the scalar relative to its dimension. However, this does not show whether an instability appeared before the near-extremal regime was reached, and so we need to resort to the numerical solutions to shed some light on this issue. In Fig.~\ref{fig:ins} we plot the dimensionless ratio of the expectation value and the relevant coupling, in terms of the normalizable and non-normalizable coefficients of the numerical solutions, for different values of $\cQ\propto \mu_r/t_r$. At the points where the ratio becomes zero there is a zero frequency quasinormal mode and most likely there will be an instability at larger values of $\cQ$. In all cases the plots show that the onset of the instability ventures in the nonextremal region only for values of the charge of the scalar $q$ that are above the bound \eqref{eq:dimbound}.  Therefore, as long as the values of the dimension and the charge are such that the near-extremal solution is stable, we do not expect the solution to be unstable against condensation. Keeping in mind the possibility of having other instabilities, we conclude that the speed of sound bound can be violated in physical states at large densities.

\section{Speed of sound in $R$-charged states}\label{sec:Rcharge}

In the previous section, we found a family of models for which the bound on the speed of sound was violated. Our approach in this was bottom up; we considered a generic Einstein-Maxwell action with a minimally coupled charged scalar, but we did not try to embed it in any string theory construction. A natural concern is--whether the results we obtained also apply in a {\em bona fide} gravity dual, or whether the additional structures in string theory will always restrict parameters in such a way that the bound is preserved. 

In order to answer this question we will study a closely related string theory model. The field theory is $\cN=4$ $SU(N_c)$ super–Yang-Mills at nonzero $R$-charge density. There are a variety of charged states that have a holographic dual description in terms of rotating branes and giant gravitons \cite{Cvetic:1999xp,McGreevy:2000cw,Myers:2001aq} in an asymptotically AdS$_5\times S^5$ space. A subclass is captured by the consistent truncation of five-dimensional $\cN=8$ SUGRA found in \cite{Gunaydin:1985cu} (the truncation is nicely presented in \cite{Bobev:2010de}, which we follow closely). There are three $U(1)$ gauge fields $A_\mu^{(I)}$, $I=1,2,3$ corresponding to the three commuting $R$-charges in the Cartan subgroup of the $SU(4)$ $R$-symmetry group. There are also four complex scalars $\zeta_i=\tanh(\varphi_i) e^{i\theta_i}$, $i=1,2,3,4$ dual to the gaugino bilinears
\begin{equation} 
\varphi_i  \ \leftrightarrow  \ {\rm tr} \lambda_i\lambda_i +{\rm h.c.} \ ,
\end{equation}
and two real scalars $\alpha,\beta$ dual to the $\cN=4$ scalar bilinears
\begin{equation}
\alpha  \ \leftrightarrow \ \,{\rm tr}\,(X_1^2+X_2^2+X_3^2+X_4^4-2 X_5^2-2X_6^2)\ , \ \beta \ \leftrightarrow \ \,{\rm tr}\,(X_1^2+X_2^2-X_3^2-X_4^2) \ .
\end{equation}

For convenience, we define $\rho=e^\alpha$ and $\nu=e^\beta$, whereby the Lagrangian density becomes
\bea
e^{-1}\cL  & = &\frac{1}{4}R-\frac{1}{4g^2}\left[\rho^4 \nu^{-4} F_{\mu\nu}^{(1)}F^{(1)\, \mu\nu}+\rho^4 \nu^{4} F_{\mu\nu}^{(2)}F^{(2)\, \mu\nu}+\rho^{-8}  F_{\mu\nu}^{(3)}F^{(3)\, \mu\nu}\right] \\
&& +\frac{1}{2}\sum_{i=1}^4 (\partial_\mu\varphi_i)^2+3(\partial_\mu\alpha)^2+(\partial_\mu \beta)^2 +\frac{1}{8}\sum_i \sinh^2(2\varphi_i)\left(\partial_\mu \theta_i+c_{iI} A_\mu^{(I)}\right)^2-V \nonumber\ ,
\eea
where
\begin{equation}
c_{iI}=\left( 
\begin{array}{ccc}
1 & 1 & -1\\
1 & -1 & 1\\
-1 & 1 & 1\\
-1 & -1 & -1
\end{array}
\right).
\end{equation}
The scalar potential is determined by a superpotential $W$ as follows \cite{Khavaev:2000gb}
\begin{equation}
V=\frac{g^2}{8}\left[\frac{1}{6}\left(\frac{\partial W}{\partial\alpha} \right)^2 +\frac{1}{2}\left(\frac{\partial W}{\partial\beta} \right)^2+\sum_{i=1}^4 \left(\frac{\partial W}{\partial\varphi_i} \right)^2  \right]-\frac{g^2}{3}W^2,
\end{equation}
where
\bea
W & = & -\frac{1}{4\rho^2\nu^2}\left[(1+\nu^4-\nu^2\rho^6) \cosh(2\varphi_1)+(-1+\nu^4+\nu^2\rho^6) \cosh(2\varphi_2)\right. \nonumber\\
&& \left.+(1-\nu^4\nu^2\rho^6) \cos(2\varphi_3)+(1+\nu^4+\nu^2\rho^6) \cos(2\varphi_4) \right]\ .
\eea

When the three charges are equal, $A_\mu^{(I)}=2 A_\mu/\sqrt{3}$, and the scalars are turned off,  the SUGRA action reduces to Einstein-Maxwell with a cosmological constant and AdS Reissner-Nordstr\"om is a solution to the equations of motion \cite{Romans:1991nq,Chamblin:1999tk} (see also \cite{Bobev:2010de} for more general charged black hole solutions).

If we take the scalars into account but remain in the equal charge case, then it is consistent with the equations of motion to set $\alpha=\beta=0$. By setting $\varphi_i=\varphi/(2\sqrt{2})$ and 
\begin{equation}
\theta_1=\theta_2=\theta_3=-\theta/\sqrt{3} \ , \ \theta_4=\sqrt{3}\theta\ ,
\end{equation}
the action becomes
\begin{equation}
e^{-1}\cL =\frac{1}{4}R-\frac{1}{g^2}F_{\mu\nu} F^{\mu\nu}+ \frac{1}{4}(\partial_\mu\varphi)^2+\frac{1}{2} \sinh^2\left(\frac{\varphi}{\sqrt{2}}\right)\left(\partial_\mu \theta- 2A_\mu\right)^2-V_\varphi \ ,
\end{equation}
where 
\begin{equation}
V_\varphi=-\frac{3g^2}{16}\left(3+\cosh(\sqrt{2}\varphi)\right)\ .
\end{equation}
The coupling constant $g$ is related to the AdS radius as $g=2/L$. If we expand to quadratic order in $\varphi$, we get that the scalar terms of the action are
\begin{equation}
e^{-1}\cL_\varphi \simeq \frac{1}{4}\left[(\partial_\mu\varphi)^2+\varphi^2(\partial_\mu \theta-2A_\mu)^2-\frac{12}{L^2}-\frac{3}{L^2}\varphi^2\right]=\frac{1}{4}\left[\left|D_\mu \phi\right|^2 -\frac{12}{L^2}-\frac{3}{L^2}|\phi|^2\right] \ .
\end{equation}
In this Lagrangian we have defined the complex scalar field $\phi=\varphi e^{i\theta}$ and the covariant derivative $D_\mu\phi=\partial_\mu\phi-2 i A_\mu\phi$. Therefore, for a small amplitude of the scalar field, the dynamics reduce to those of a field of charge $q=2$ and mass $m^2L^2=-3$, dual to a $\Delta=3$ scalar operator which is a combination of components of the gaugino bilinears. The coefficient of the quartic term as in \eqref{eq:V3} is $V_4=-1$. This is within the class of models we are considering, but the dimension is integer, and the relation between the dimension and the charge does not satisfy the bound \eqref{eq:dimbound}. This introduces some technical complications, as the asymptotic expansion of the scalar close to the AdS boundary is modified by the introduction of logarithmic terms
\begin{equation}\label{eq:rexp2}
\phi \simeq \frac{L^2}{r} \widetilde{\phi}_{(0,0)}+\frac{L^6}{r^3}\widetilde{\phi}_{(2,1)}\log\frac{r}{L}+\left(\frac{L}{r}\right)^{3} L^3 \phi_{(0,0)}.
\end{equation}
We will redo the analysis for this special case in the following.

\subsection{Solutions}

The equations of motion for the scalar field can be obtained from \eqref{eq:phiu} by setting $\nu=1$ in \eqref{eq:zetacoeffs}. The leading terms in the expansion close to the AdS boundary are
\begin{equation}
\phi\simeq \alpha_- u^{1/2}+ \tilde{\alpha}_+ u^{3/2}\log u +\alpha_+u^{3/2} \ ,
\end{equation}
where $\tilde{\alpha}_+$ is determined by $\alpha_-$.  We can find an analytic near-extremal solution following the same procedure discussed in the previous section, although the details are slightly different because the dimension of the scalar operator is an integer number. The full calculation can be found in Appendix~\ref{app:nearextremal2}. The ratio between normalizable and non-normalizable coefficients takes the same form as in the previous examples \eqref{eq:ratioapam}, but in this case $\lambda=i/\sqrt{2}$ is complex [the bound \eqref{eq:dimbound} is not satisfied]. To go beyond the near-extremal limit, we use numerics, and the solutions are computed in the same way as in the previous section.

Once we have obtained the numerical solution we extract the values of the coefficients of the normalizable and non-normalizable modes by evaluating the solution at the boundary cutoff $u=\epsilon_0$. For $\nu=1$, we define the numerical values as
\bea
(\alpha_-)_n & = & \left. u^{-1/2} \phi_n(u) \right|_{u=\epsilon_0} \\
(\tilde{\alpha}_+)_n & = & \left.  u \partial_u^2\left[ u^{-1/2} \phi_n(u)\right] \right|_{u=\epsilon_0} \\
(\alpha_+)_n & = & \left. \partial_u\left[ u^{-1/2} \phi_n(u)\right] -(\tilde{\alpha}_+)_n(\log u+1) \right|_{u=\epsilon_0}\ .
\eea
These values are determined by $\cQ$ and $q$. In the numerical calculation, we have fixed the amplitude of the scalar field. Since the equations of motion are linear, we can again generate a full set of values by doing a trivial rescaling,
\begin{equation}
\alpha_-=a (\alpha_-)_n \ , \ \tilde{\alpha}_+=a(\tilde{\alpha}_+)_n \ , \  \alpha_+=a(\alpha_+)_n  
\end{equation}
for some real number $a$. We determine the value of $a$ by fixing the temperature in units of the relevant coupling \eqref{eq:reducedmut2} for each value of $\cQ$.

\subsection{Thermodynamics }

We have computed the free energy density in \eqref{eq:freeenergy2}, in terms of the coefficients of the asymptotic expansions given in \eqref{eq:nbseries2}. Comparing with the pressure $p=\vev{T^{ii}}$, given in \eqref{eq:tii2}, we see that it is equal to minus the pressure $\cF=-p$. The expressions for the energy density $\varepsilon=\vev{T^{00}}$ and the charge density $n=\vev{J^0}$ can be found in \eqref{eq:t002} and \eqref{eq:j02}, respectively. Using these expressions as well as \eqref{eq:solsbge2} and \eqref{eq:betaH}, one finds the usual thermodynamic relation
\begin{equation}
\varepsilon+p=\mu n+Ts \ .
\end{equation}
We again introduce the reduced quantities \eqref{eq:reducedmut}  and \eqref{eq:reduced} with $\n=1$; in particular, the reduced temperature and chemical potential read
\be \label{eq:reducedmut2}
\m_r \equiv \frac{\mu}{J}= \frac{\sqrt{3}}{2} \frac{\cQ}{\alpha_-} \ , \ t_r\equiv\frac{T}{J} = \frac{2-\cQ^2}{2\pi \alpha_-} \ .
\ee
The energy density and pressure depend on the enthalpy $w=\varepsilon+p$ and the expectation value of the scalar operator as
\bea
\varepsilon_r & = & \frac{3}{4}w_r  +\frac{1}{2}  \tilde{v}_r \\
p_r & = & \frac{1}{4}w_r -\frac{1}{2}  \tilde{v}_r \\
\tilde{v}_r & = & v_r -\frac{1}{2} q^2\mu_r^2 +\left(\frac{1}{6}+\frac{V_4}{4}\right)\ .
\eea
Comparing with the Reissner-Nordstr\"om solution for the background metric \eqref{eq:holomet} with the expansions \eqref{eq:nbseries2} and using the expressions for the renormalized values \eqref{eq:t003} and \eqref{eq:tii3} and the relation with the coefficients in the $u$ expansion \eqref{eq:mapru}, the leading order contributions are
\bea
w_r & = & 4 \frac{1}{\alpha_-^4}(1+\cQ^2) +O\left( \frac{\log \alpha_-}{\alpha_-^2}\right) \\
\tilde{v}_r & = & -2\frac{\alpha_+}{\alpha_-}\frac{1}{\alpha_-^2} +q^2\mu_r^2\log\alpha_-+\kappa_1 q^2 \mu_r^2+O(\log\alpha_-) \ .
\eea

\subsection{Speed of sound and stability}

Through \eqref{eq:reducedmut2}, an isothermal variation will satisfy
\begin{equation}
\alpha_-(\cQ)=\frac{1}{\pi t_r} \left(1-\frac{\cQ^2}{2}\right) \ .
\end{equation}
For asymptotically large values of the chemical potential, the changes in enthalpy and the term associated to the breaking of conformal invariance are, to leading order,
\bea
dw_r & \simeq  & 24\frac{\cQ}{\alpha_-^4} \frac{2+\cQ^2 }{2-\cQ^2}d\cQ \\
d\tilde{v}_r & \simeq & \frac{3}{2} q^2\cQ \frac{2+\cQ^2}{2-\cQ^2} \frac{\log\alpha_-}{\alpha_-^2}d\cQ \ ,
\eea
giving
\begin{equation}
\frac{d v_r}{d w_r} \simeq  \frac{q^2}{16} \alpha_-^2\log\alpha_-\ .
\end{equation}
Similarly, the isothermal speed of sound is
\begin{equation}
v_s^2=\left( \frac{\partial p_r}{\partial \epsilon_r}\right)_{t_r} \simeq \frac{1}{3}\left(1-\frac{8}{3} \frac{d v_r}{d w_r}\right) \ .
\end{equation}

\begin{figure}
	\begin{center}
		\includegraphics[width=0.75\textwidth]{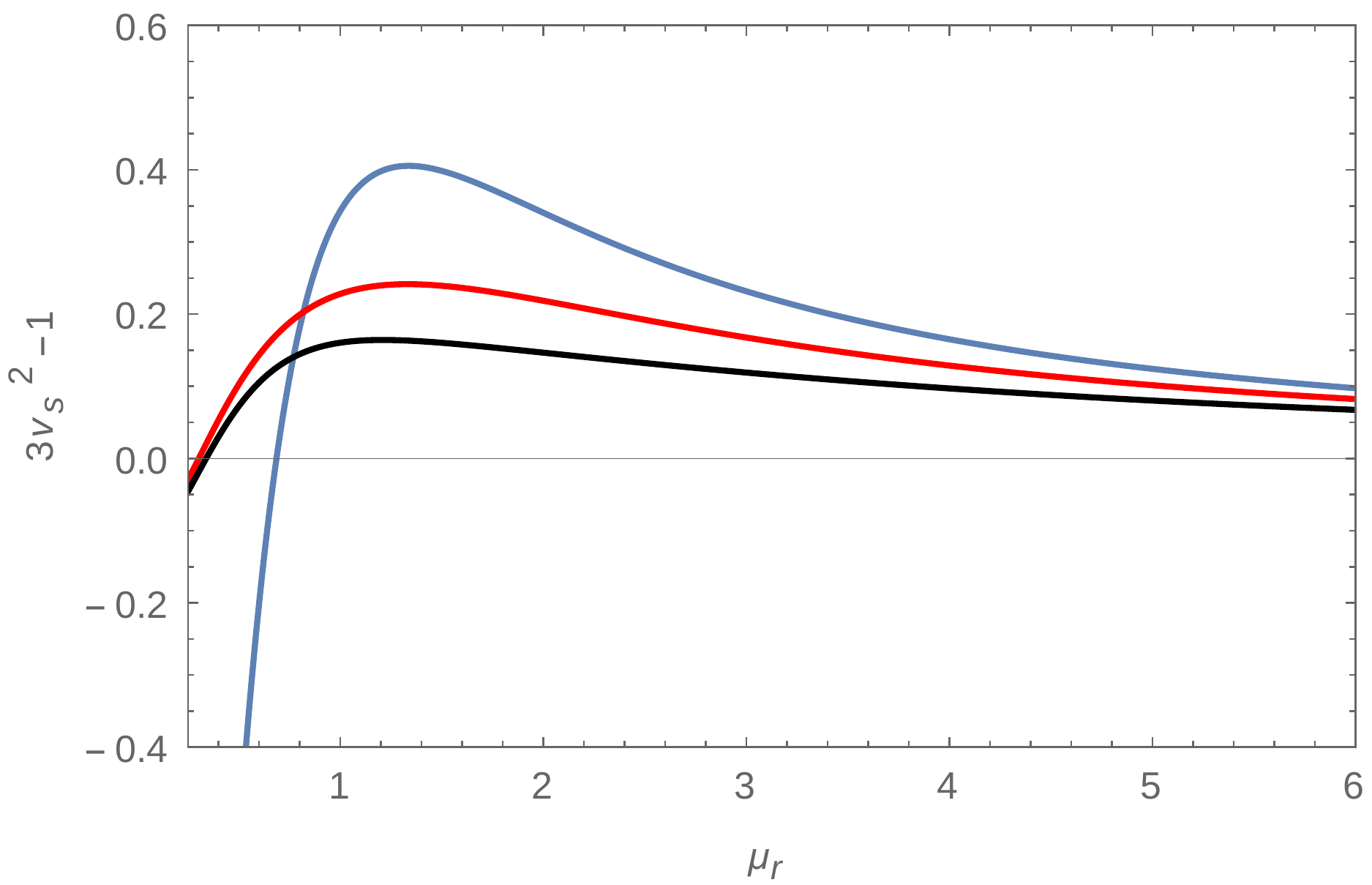}
		\end{center}
		\caption{$3v_s^2-1$ as a function of the chemical potential for different values of the temperature: $t_r=0.1$ (blue), $t_r=0.5$ (red), and $t_r=1$ (black). Notice that for all values of the temperature,  the curves approach the conformal value at large values of the chemical potential.}\label{fig:vsD3}
\end{figure}

At large values of the chemical potential the solution is near-extremal $\cQ\simeq \sqrt{2}$ and 
\begin{equation}
\alpha_-^2\simeq \frac{3}{2 \mu_r^2} \ll 1\, ,
\end{equation}
while the speed of sound becomes
\begin{equation}
v_s^2\simeq \frac{1}{3}\left(1+ \frac{1}{4}q^2  \frac{\log \mu_r}{\mu_r^2}\right).
\end{equation}
The speed of sound will approach the conformal value from above at asymptotically large densities. However, we expect to have instabilities in the extremal limit, and in order to go beyond this limit we will resort to numerics. The result will depend on the choice of a finite counterterm $\kappa_1$ that is shown explicitly in \eqref{eq:t003} and \eqref{eq:tii3}. We will set $\kappa_1=0$, noting that as long as this parameter is not large, it will not qualitatively affect to the results. We present the deviation of the isothermal speed of sound from the conformal value, $3 v_s^2-1$, in Fig.~\ref{fig:vsD3} for different values of the temperature. We see that the speed of sound remains above the conformal value until relatively low values of the chemical potential. The value $\mu_r=1$ corresponds to $\cQ\simeq  0.83 \sqrt{2}, 0.43 \sqrt{2}, 0.24 \sqrt{2}$ for $t_r=0.1,0.5,1$ respectively.

As we found in the previous section (see Fig.~\ref{fig:ins} for $\nu=1$), we expect the instability that will trigger the formation of a homogeneous condensate to appear close to the near-extremal regime, since we found a critical charge of $q^*\simeq 1.47$ for $\cQ=\sqrt{2}-10^{-3}$,  which is close to the value we have in the $R$-charged solution $q=2$. For $q=2$, we find that the instability appears at $\cQ^*\simeq 1.396$ (purple curve in Fig.~\ref{fig:ins}). This corresponds to values of the chemical potential $\mu_r\simeq 14.8, 74.2, 148.4$ for $t_r=0.1,0.5,1$, respectively. We have thus shown that the speed of sound can be violated in the $R$-charged state with no obvious instabilities.

\section{Conclusions and outlook}\label{sec:conclusions}

According to current understanding, it appears likely that the speed of sound in neutron star matter exceeds the conformal value $v_s=1/\sqrt{3}$ in the dense nuclear matter phase \cite{Bedaque:2014sqa}. Knowing that for ultradense quark matter, the speed of sound approaches this value from below, we are in practice left with two possibilities: either the quantity exhibits a discontinuous jump in a first order phase transition or it must continuously first decrease and then increase its value in the quark matter phase. The latter scenario has, however, been disfavored due to the lack of first principles calculations exhibiting speeds of sound larger than the conformal value in deconfined matter. In addition to perturbative calculations, this statement holds true for all known holographic setups that flow to a four-dimensional CFT in the UV, which has prompted speculation of a more fundamental speed of sound bound \cite{Hohler:2009tv,Cherman:2009tw}. 

In the paper at hand, we have shown that the conjectured bound on the speed of sound in holographic models with UV fixed points is violated for a simple class of models involving RG flows triggered by relevant scalar operators charged under a global Abelian symmetry. Within this class, we were able to find a string theory example: a charged black hole dual to $\cN=4$ theory at finite $R$-charge density, deformed by a gaugino mass term. Since at very large densities an instability toward the formation of a homogeneous condensate will develop, we made sure that the violation occurs in the stable regime. We may conclude that \textit{there is no universal bound for the speed of sound in holographic models dual to ordinary four-dimensional relativistic field theories}. This comes as good news for everyone wishing to build realistic holographic models for high-density nuclear or quark matter.

A natural extension of our work is clearly to go beyond the approximation of a small breaking of conformal invariance and study how high values the speed of sound can be maximally obtained. Of particular interest is to investigate whether large enough speeds can be obtained that would allow the building of stable hybrid stars with holographic quark matter in their cores. Continuing along these lines, it might be interesting to study the behavior of bottom-up models designed to match the properties of QCD at zero \cite{Gursoy:2008bu,Gursoy:2009jd,DeWolfe:2010he,Rougemont:2015wca,Rougemont:2015ona} and finite density \cite{Alho:2013hsa} and to see if they give phenomenologically sensible results at very small temperatures.

A different but equally interesting challenge to pursue would be to find a top-down model with finite baryon (rather than $R$-charge) density. Quark matter is typically introduced by embedding probe branes in the geometry, and in the known examples where the theory is truly ($3+1$)-dimensional the bound is satisfied even at finite density. This might change upon considering the backreaction of the branes. Solutions with backreacted flavors at finite density have been recently constructed in \cite{Bigazzi:2011it,Bigazzi:2013jqa,Faedo:2014ana,Faedo:2015urf}.  Another possibility is to take an alternative large-$N$ limit where (anti) fundamental fields are extrapolated to two-index antisymmetric representations  \cite{HoyosBadajoz:2009hb} and where operators with baryon charge map to gravitational modes.  We leave these investigations for future work.

\vspace{5mm}


\paragraph{Acknowledgments}
\noindent N.J.~and A.V.~are supported by the Academy of Finland Grants No.~1273545, No. 1297472, and No. 1303622, while C.H.~and D.R.F.~are partially supported by the Spanish Grant No. MINECO-16-FPA2015-63667-P. C.H.~is in addition supported by the Ramon y Cajal fellowship RYC-2012-10370, and D.R.F. is supported by the FC-15-GRUPIN-14-108 research grant from Principado de Asturias. N.J.~wishes to thank Long Island University and Stony Brook University for warm hospitality, as well as the Nordita program on Black Holes and Emergent Spacetime and the Heraklion Workshop on Theoretical Physics for providing prolific environments during a time--when this work was being finished.

\appendix

\section{Near-extremal solutions}\label{app:nearext}

\subsection{Noninteger $\Delta$}\label{app:nearextremal1}

\subsubsection{Near-horizon solution}

First we expand (\ref{eq:phiu}) in the region close to the horizon. It is convenient to introduce a new radial coordinate $v$, defined as
\begin{equation}
u=1-\frac{4}{3}\epsilon v \ .
\end{equation}
We then expand (\ref{eq:phiu}) to leading order in $\epsilon$. In this new coordinate, the horizon is located at $v=0$ and the asymptotic region gets pushed to $v> 1/\epsilon\to \infty$. 
The solution to the equation of motion that is regular as $v\to 0$ is
\begin{equation}
\phi=\phi_H (1+v)^{-i\frac{q}{2\sqrt{6}}} {}_2 F_1\left[\frac{1}{2}-i\frac{q}{2\sqrt{6}}-\frac{\lambda}{2\sqrt{3}} \ , \ \frac{1}{2}-i\frac{q}{2\sqrt{6}}+\frac{\lambda}{2\sqrt{3}} \,,\, 1\,,\, -v \right]+ c.c.\ ,
\end{equation}
where we have defined $\lambda= \sqrt{\nu^2-\frac{q^2}{2}-1}$ and where $\phi(0)=\phi_H$ is the value of the scalar field at the horizon. The expansion as $v\to \infty$ gives, to leading order
\begin{equation}\label{horexp}
\phi \sim \phi_H \left( C_- v^{-\frac{1}{2} -\frac{\lambda}{2\sqrt{3}}}+C_+ v^{-\frac{1}{2} +\frac{\lambda}{2\sqrt{3}}}\right) \ ,
\end{equation}
where
\begin{equation}
C_\pm = \frac{\Gamma\left(\pm \frac{\lambda}{\sqrt{3}} \right)}{\left|\Gamma\left(\frac{1}{2}-i\frac{q}{2\sqrt{6}}\pm \frac{\lambda}{2\sqrt{3}} \right) \right|^2} \ .
\end{equation}

\subsubsection{Asymptotic solution}

We can find the leading order solution simply setting $\epsilon\to 0$ in the equation, giving the solution in the extremal Reissner-Nordstr\"om geometry
\begin{equation}
\phi= \alpha_+ \varphi_++\alpha_-\varphi_- \ ,
\end{equation}
where $\alpha_\pm$ are arbitrary coefficients and
\bea
\varphi_\pm & = & u^{1\pm \frac{\nu}{2}} (1-u)^{-\frac{1}{2}-\frac{|q-\lambda |}{q-\eta} \frac{\lambda}{2\sqrt{3}}  } (1+2 u)^{-\frac{1\pm\nu}{2} +\frac{|q-\lambda |}{q-\lambda} \frac{\lambda}{2\sqrt{3}}  }    \\
& & \times {}_2 F_1\left[\frac{1\pm \nu}{2}+\frac{|q-\lambda|}{2\sqrt{3}}\,,\, \frac{1\pm \nu}{2}-\frac{q+\lambda}{q-\lambda}\frac{|q-\lambda|}{2\sqrt{3}} \,,\, 1\pm\nu \,,\, \frac{3u}{1+2 u} \right] \ .
\eea
Expanding close to the boundary  $u\to 0$, the coefficients behave as
\begin{equation}
\varphi_\pm \sim u^{1\pm\frac{\nu}{2}} \ .
\end{equation}
Therefore, $\varphi_+$ is the normalizable mode, and $\varphi_-$ is the non-normalizable mode. We now evaluate the solution at $u=1-\frac{4}{3}\epsilon v$ and expand for $\epsilon \to 0$. The expansion takes the same form as for the near-horizon solution
\begin{equation}\label{asympexp}
\varphi_\pm \sim\left( D_{\pm-} v^{-\frac{1}{2} -\frac{\lambda}{2\sqrt{3}}}+D_{\pm+} v^{-\frac{1}{2} +\frac{\lambda}{2\sqrt{3}}}\right)\ .
\end{equation}
The coefficients and exponents depend on the sign of $q-\lambda$, but the final result is the same for both $q>\lambda $ and $\lambda<q$.

\subsubsection{Matching}

We have to match the coefficients of the near-horizon and asymptotic solutions in such a way that the leading order terms in the overlapping region \eqref{horexp} and \eqref{asympexp} are the same. This gives the conditions
\begin{equation}
 \alpha_+ D_{\pm +}+\alpha_- D_{\pm -}=\phi_H C_\pm\ , 
\end{equation} 
thus fixing $\alpha_+$ and $\alpha_-$ in terms of $\phi_H$ and the coefficients of the expansion. In the end we should fix the coefficient of the non-normalizable mode to be equal to the source of the dual operator. That fixes the value of the scalar at the horizon and the coefficient of the normalizable mode, both of which are also proportional to the source. The proportionality coefficient of the normalizable mode that determines the VEV of the dual operator is proportional to
\begin{equation}\label{eq:cnu1}
\frac{\alpha_+}{\alpha_-}=-3^{\nu }\frac{ \Gamma (1-\nu ) \Gamma \left(\frac{1+\nu}{2}+\frac{\lambda+q}{2\sqrt{3}} \right) \Gamma \left(\frac{1+\nu}{2}+\frac{\lambda-q}{2\sqrt{3}}  \right)}{\Gamma (1+\nu) \Gamma \left(\frac{1-\nu}{2}+\frac{\lambda+q}{2\sqrt{3}} \right) \Gamma \left(\frac{1-\nu}{2}+\frac{\lambda-q}{2\sqrt{3}}  \right)}\frac{1+\beta_1\left(\frac{4\epsilon}{9}\right)^{\lambda/\sqrt{3} } }{1+\beta_2\left(\frac{4\epsilon}{9}\right)^{\lambda/\sqrt{3} } } \ ,
\end{equation}
where 
\begin{equation}
\beta_1= \frac{\Gamma\left( -\frac{\lambda}{\sqrt{3}}\right)\Gamma\left(1 -\frac{\lambda}{\sqrt{3}} \right) }{\Gamma\left( \frac{\lambda}{\sqrt{3}} \right)\Gamma\left(1 +\frac{\lambda}{\sqrt{3}} \right)}\frac{\Gamma\left(\frac{1}{2} -i\frac{q}{2\sqrt{6}} +\frac{\lambda}{2\sqrt{3}} \right)\Gamma\left( \frac{1}{2} +i\frac{q}{2\sqrt{6}} +\frac{\lambda}{2\sqrt{3}} \right) }{\Gamma\left( \frac{1}{2} -i\frac{q}{2\sqrt{6}} -\frac{\lambda}{2\sqrt{3}} \right)\Gamma\left( \frac{1}{2} +i\frac{q}{2\sqrt{6}} -\frac{\lambda}{2\sqrt{3}} \right) }\frac{\Gamma\left(\frac{1-\nu}{2}+\frac{\lambda+q}{2\sqrt{3}} \right)\Gamma\left( \frac{1-\nu}{2}+\frac{\lambda-q}{2\sqrt{3}} \right)}{\Gamma\left( \frac{1-\nu}{2}-\frac{\lambda+q}{2\sqrt{3}} \right)\Gamma\left( \frac{1-\nu}{2}-\frac{\lambda-q}{2\sqrt{3}} \right)}
\end{equation}
and
\begin{equation}
\beta_2=\frac{\cos\left(\frac{q\pi}{\sqrt{3}} \right)+\cos\left[\pi\left( \frac{\lambda}{\sqrt{3}}-\nu\right)\right] }{\cos\left(\frac{q\pi}{\sqrt{3}} \right)+\cos\left[\pi\left( \frac{\lambda}{\sqrt{3}}+\nu\right)\right] }\beta_1\ .
\end{equation}
The terms that go as $\sim \epsilon^{\lambda/\sqrt{3}}$ can be neglected only if $\lambda>0$. This requires
\begin{equation}
\nu^2>1+\frac{q^2}{2} \ \ \Rightarrow \ \ \Delta > 2+\sqrt{1+\frac{q^2}{2}} \ .
\end{equation}
If $q^2\geq 6$ this is not possible for $\Delta < 4$.

\subsection{Solutions for $\Delta=3$}\label{app:nearextremal2}

We can set $\nu=1$ in Eq. \eqref{eq:zetacoeffs} and follow the same procedure.

\subsubsection{Near-horizon solution}

We can convert the equation of motion in a hypergeometric equation by extracting a factor from the scalar field $\phi=(1+v)^{-i \frac{q}{2\sqrt{6}}} X$. The resulting equation is (we keep $q$ as a parameter)
\begin{equation}
v(1+v) X''+\left( 1+\left(2-i\frac{q}{\sqrt{6}}\right)v\right)X'+\left(\frac{1}{4}-i\frac{q}{2\sqrt{6}} \right)X=0 \ .
\end{equation}
The regular solution is
\begin{equation}
X={}_2F_1\left(\frac{1}{2}\,,\,\frac{1}{2}-i\frac{q}{\sqrt{6}}\,,\, 1\,,\, -v \right) \ .
\end{equation}
In order to make the solution real we combine it with the complex conjugate:
\begin{equation}
\phi=\frac{\phi_H}{2}\left[ (1+v)^{-i \frac{q}{2\sqrt{6}}}  {}_2F_1\left(\frac{1}{2}\,,\,\frac{1}{2}-i\frac{q}{\sqrt{6}}\,,\, 1\,,\, -v \right)+(1+v)^{i \frac{q}{2\sqrt{6}}}  {}_2F_1\left(\frac{1}{2}\,,\,\frac{1}{2}+i\frac{q}{\sqrt{6}}\,,\, 1\,,\, -v \right)\right] \ .
\end{equation}
The expansion as $v\to\infty$ is
\begin{equation}\label{horexp2}
\phi \sim \frac{\phi_H}{2}\left( C_-v^{-\frac{1}{2}-i\frac{q}{2\sqrt{6}} }+C_+v^{-\frac{1}{2}+i\frac{q}{2\sqrt{6}} }\right) \ ,
\end{equation}
where
\begin{equation}
C_\pm=\frac{\Gamma\left(\pm i\frac{q}{\sqrt{6}} \right)}{\sqrt{\pi}\Gamma\left(\frac{1}{2}\pm i\frac{q}{\sqrt{6}} \right)} \ .
\end{equation}

\subsubsection{Asymptotic solution}

We can find the leading order solution simply setting $\epsilon\to 0$ in (\ref{eq:phiu}): 
\begin{equation}\label{eq:asphiudeltathree}
\phi''-\frac{1+u+4 u^2}{u(1-u)(2u+1)}\phi'+\frac{3}{8}\frac{2+(4+q^2) u}{u^2(1-u)^2(1+2u)^2}\phi=0\ .
\end{equation}
We can convert this in a hypergeometric equation by defining
\begin{equation}
\hat{\varphi}_\mp=(2u+1)^{\pm i\frac{q}{2\sqrt{6}}} (1-u)^{-\frac{1}{2}\mp i\frac{q}{2\sqrt{6}}} u^{\frac{1}{2}} X_\mp\left[ \frac{3 u}{2u+1}\right]\ ,
\end{equation}
and changing variables to 
\begin{equation}
z=\frac{3 u}{2u+1}, \ \ u=\frac{z}{2z-3}\ .
\end{equation}
Equation (\ref{eq:asphiudeltathree}) becomes
\begin{equation}
z(1-z) X_\mp''+\left(\pm i\frac{q}{\sqrt{6}}\right)z X_\mp'+\frac{q^2}{8}X_\mp=0\ .
\end{equation}
The two independent solutions are
\begin{equation}
X_\mp={}_2 F_1\left( \frac{q}{2\sqrt{3}} \left(1\mp\frac{i}{\sqrt{2}}\right)\,,\,  -\frac{q}{2\sqrt{3}} \left(1\pm\frac{i}{\sqrt{2}}\right)\,,\, 1\mp  i\frac{q}{\sqrt{6}}\,,\, 1-z\right)\ .
\end{equation}
The general scalar solution is thus
\begin{equation}
\phi= \hat{\alpha}_+ \hat{\varphi}_++\hat{\alpha}_-\hat{\varphi}_-\ ,
\end{equation}
where $\hat{\alpha}_\pm$ are arbitrary coefficients. 

Expanding close to the boundary  $u\to 0$
\begin{equation}
\hat{\varphi}_\mp \sim A_\mp u^{1/2}+B_\mp u^{3/2}\ .
\end{equation}
The coefficients $A_\mp$ correspond to the non-normalizable mode, and the coefficients $B_\mp$ correspond to the normalizable mode. We define the non-normalizable $\varphi_-$ and normalizable $\varphi_+$ modes as
\bea
\varphi_- & = & \frac{B_-\hat{\varphi}_+- B_+\hat{\varphi}_-}{B_-A_+-B_+A_-} \\
\varphi_+ & = & \frac{A_-\hat{\varphi}_+- A_+\hat{\varphi}_-}{A_-B_+-A_+B_-} \ .
\eea
The general solution is
\begin{equation}
\varphi=\alpha_+\varphi_++\alpha_-\varphi_- \ .
\end{equation}
We now evaluate the solution at $u=1-\frac{4}{3}\epsilon v$ and expand for $\epsilon \to 0$. The expansion takes the same form as for the near-horizon solution
\begin{equation}\label{asympexp2}
\varphi_\pm \sim\left( D_{\pm -}v^{-\frac{1}{2}-i\frac{q}{2\sqrt{6}} }+D_{\pm +}v^{-\frac{1}{2}+i\frac{q}{2\sqrt{6}} }\right)\ .
\end{equation}

\subsubsection{Matching}

We have to match the coefficients of the near-horizon and asymptotic solutions in such a way that the leading order terms in the overlapping region \eqref{horexp2} and \eqref{asympexp2} are the same. This gives the conditions
\begin{equation}
 \alpha_- D_{\pm -}+\alpha_+ D_{\pm +}=\phi_H C_\pm \ . 
\end{equation} 
This fixes $\alpha_-$ and $\alpha_+$ in terms of $\phi_H$ and the coefficients of the expansion. In the end, we should fix the coefficient of the non-normalizable mode to be equal to the source of the dual operator. That fixes the value of the scalar at the horizon and the coefficient of the normalizable mode, both of which are also proportional to the source. The proportionality coefficient of the normalizable mode that determines the VEV of the dual operator is proportional to
\begin{equation}\label{eq:c1}
\frac{\alpha_+}{\alpha_-}=\left[\frac{1}{2} -i\frac{\sqrt{6}q }{4}-\frac{3}{8}q^2 \left(H_{\frac{2+i\sqrt{2} }{4\sqrt{3}}q }+H_{\frac{-2+i\sqrt{2} }{4\sqrt{3}}q }+\log 3 \right) \right]\frac{1+\beta_1\left(\frac{4\epsilon}{9}\right)^{iq/\sqrt{6} } }{1+\beta_2\left(\frac{4\epsilon}{9}\right)^{iq/\sqrt{6} } }\ ,
\end{equation}
where 
\begin{equation}
\beta_2=\frac{-3+i\sqrt{6}q}{q^2\Gamma(q c)} \left[\frac{\Gamma\left( 1-i\frac{q}{\sqrt{6}} \right) }{\Gamma\left( i\frac{q}{\sqrt{6}} \right)}\right]^2 \frac{\Gamma\left(-\frac{1}{2}+i\frac{q}{\sqrt{6}}  \right)}{\Gamma\left(\frac{1}{2}-i\frac{q}{\sqrt{6}}  \right)} \frac{\Gamma\left(\frac{2+i\sqrt{2} }{4\sqrt{3}}q \right)\Gamma\left(\frac{-2+i\sqrt{2} }{4\sqrt{3}}q \right)}{\Gamma\left(\frac{2-i\sqrt{2} }{4\sqrt{3}}q  \right)} \ .
\end{equation}
Here
\begin{equation}
c=-\frac{\sqrt{1+i2\sqrt{2}}}{2\sqrt{6}}\simeq -0.29-0.2 i \ .
\end{equation}
The other coefficient is
\begin{equation}
\beta_1=\frac{1+i\sqrt{\frac{3}{2}}q-\frac{3}{4}q^2 \left(-1+H_{\frac{2-i\sqrt{2}}{4\sqrt{3}}q }+H_{cq} +\log 3 \right) }{1-i\sqrt{\frac{3}{2}}q-\frac{3}{4}q^2 \left(-1+H_{\frac{2+i\sqrt{2}}{4\sqrt{3}}q }+H_{\frac{-2+i\sqrt{2}}{4\sqrt{3}}q } +\log 3 \right) }\beta_2 \ .
\end{equation}
Note that the terms that go as $\sim \epsilon^{iq/\sqrt{6}}$ cannot be neglected.

\section{Useful formulas}\label{app:formulas}

\subsection{Equations of motion and boundary expansions}

In the presence of a nonzero scalar $\phi(r)$ and gauge field $A_0(r)$, we consider an ansatz for the metric of the form
\begin{equation}
ds^2=\frac{L^2}{r^2 f(r)}dr^2+\frac{r^2}{L^2}e^{2A(r)}\left(-f(r)dt^2+dx^2\right) \ .
\end{equation}
The equations of motion  \eqref{EOM} become
\bea
\label{eqf} 0 &=&f''+\left( 4A'+\frac{5}{r} \right)f'-L^4 e^{-2A}\left[\frac{4}{r^2}A_0'^2+ \frac{2q^2}{r^4 f}A_0^2\phi^2  \right] \\
\label{eqA0} 0 &=&A_0''+\left( 2A'+\frac{3}{r}\right)A_0'- q^2 \frac{\phi^2}{2r^2f}A_0 \\
0 &=&A''+\frac{1}{r}A'+\frac{1}{3}\left(\phi'^2+L^4 q^2 e^{-2A}\frac{A_0^2 \phi^2}{r^4 f^2}\right) \\
0 &=&\phi''+\left(4A'+\frac{f'}{f}+\frac{5}{r}\right)\phi'+\left[L^4 q^2e^{-2A}\frac{A_0^2}{r^4 f^2}-\frac{\Delta(\Delta-4)}{r^2 f}\right]\phi \\
0 &=& L^4e^{-2A}\frac{A_0'{}^2}{r^2 f}+A'\left(\frac{12}{r}+\frac{3}{2}\frac{f'}{f} + 6A'\right)+\frac{6}{r^2}\left( 1-\frac{1}{f}\right)+\frac{3f'}{2rf}-\nonumber\\
& & \frac{1}{2}\phi'^2+\phi^2\left[ \frac{\Delta(\Delta-4)}{2 r^2 f}-L^4 q^2e^{-2A}\frac{A_0^2}{2r^4 f^2} \right]\ .
\eea
Assuming the mass of the scalar is such that there are no logarithmic terms (noninteger $\Delta$), the gravity solution admits the series expansion
\bea
f & = & 1 + \sum_{n,m} \frac{f_{(n,m)}}{r^{n+m\Delta}}\,, \quad A_0 = \m + \sum_{n,m} \frac{A_0{}_{(n,m)}}{r^{n+m\Delta}}\,,\quad A = \sum_{m,n} \frac{A_{(n,m)}}{r^{n + m\Delta} } \nonumber\\
\phi & = & \frac{L^{2(4-\Delta)}}{r^{4-\Delta}} \sum_{n,m} \frac{\widetilde{\phi}_{(n,m)}}{r^{n+m\Delta}} + \frac{L^{2\Delta}}{r^\Delta} \sum_{n,m} \frac{\phi_{(n,m)}}{r^{n+m\Delta}} \ , \label{eq:nbseries}
\eea
where we will identify $\widetilde{\phi}_{(0,0)}$ as the source and $\phi_{(0,0)}$ as the VEV, with dimensions $4-\Delta$ and $\Delta$, respectively. Using the equations of motion we fix the leading coefficients to be 
\bea 
\phi &\sim &\frac{L^{2(4-\Delta)}}{r^{4-\Delta}}\widetilde{\phi}_{(0,0)} + \frac{L^{2\Delta}}{r^{\Delta}}\phi_{(0,0)}+ \frac{L^{6(4-\Delta)}}{r^{3(4-\Delta)}}\frac{\Delta-4}{12(\Delta-3)}\widetilde{\phi}_{(0,0)}^3  +\frac{L^{2(6-\Delta)}}{r^{6-\Delta}}\frac{\mu^2 q^2}{4(\Delta-3)}  \widetilde{\phi}_{(0,0)}+ \ldots \nonumber\\
A &\sim & -\frac{1}{12} \frac{L^{4(4-\Delta)}}{r^{2(4-\Delta)}}\widetilde{\phi}_{(0,0)}^2 -\frac{L^{8(4-\Delta)}}{r^{4(4-\Delta)}}\frac{\Delta-4}{96(\Delta-3)}\widetilde{\phi}_{(0,0)}^4-\frac{L^{5(4-\Delta)}}{r^{2(5-\Delta)}}\frac{\mu^2 q^2 (\Delta^2-8\Delta +18)}{24(\Delta-3)(\Delta-5)^2}\widetilde{\phi}_{(0,0)}^2 \nonumber\\
& &+\frac{L^8}{r^4}\frac{\Delta(\Delta-4)}{24}\widetilde{\phi}_{(0,0)} \phi_{(0,0)}  + \ldots \nonumber \\
f & \sim & 1 + \frac{L^8}{r^4} f_{(4,0)}+ \frac{L^{4(5-\Delta)}}{r^{2(5-\Delta)}} \frac{\mu^2 q^2}{2(\Delta-3)(\Delta-5)}\widetilde{\phi}_{(0,0)}^2  +\ldots \nonumber\\
A_0 & \sim & \m + \frac{L^4 }{r^2}A_0{}_{(2,0)}+ \frac{L^{4(4-\Delta)}}{r^{2(4-\Delta)}} \frac{\mu q^2}{8(\Delta^2-7\Delta+12)}\widetilde{\phi}_{(0,0)}^2 +\ldots \ . \label{eq:nbseries1}
\eea
For $\Delta=3$, the expansions are modified due to the presence of logarithmic terms, and the asymptotic behavior of the scalar field becomes 
\be
\phi_{r\to \infty} \sim \frac{L^2}{r}\widetilde{\phi}_{(0,0)} + \frac{L^6}{r^3} \left( \widetilde{\phi}_{(1,1)}\log \frac{r}{L} + \phi_{(0,0)} \right) + \ldots \ , 
\ee
where again, we identify $\widetilde{\phi}_{(0,0)}$ as the source and $\phi_{(0,0)}$ as the VEV. Due to the presence of the logarithmic term, which is leading, the near-boundary series expansions for the gauge field and warp factors differ this time from \Eq{eq:nbseries1},
\be 
f = 1 + \sum_{n\geq m} \frac{f_{(n,m)}}{r^n}\left[\log  \left(\frac{r}{L}\right)\right]^m\ , \ A_0 = \m + \sum_{n\geq m} \frac{A_0{}_{(n,m)}}{r^n}\left[\log  \left(\frac{r}{L}\right)\right]^m\ , \  A = \sum_{n \geq m} \frac{A_{(n,m)}}{r^n}\left[\log  \left(\frac{r}{L}\right)\right]^m\ .
\ee
We will also allow a potential that is not purely quadratic but has an expansion
\begin{equation}\label{eq:V3}
V(\Phi^\dagger \Phi)=-\frac{12}{L^2}-\frac{3}{L^2}\Phi^\dagger\Phi +\frac{V_4}{2 L^2}\left(\Phi^\dagger \Phi \right)^2+\ldots \ .
\end{equation}
Inserting these series into the equations of motion, we fix the leading coefficients to be
\bea 
\phi &\sim &\frac{L^2}{r}\widetilde{\phi}_{(0,0)} + \frac{L^6}{r^3}\left[\widetilde{\phi}_{(0,0)}\left(\frac{1 }{2}\m^2 q^2 -\left(\frac{1}{3}+\frac{V_4}{2}\right) \widetilde{\phi}_{(0,0)}^2\right) \log \frac{r}{L} + \phi_{(0,0)}\right] +\ldots \nonumber\\
A &\sim &-\frac{L^4}{r^2}\frac{\widetilde{\phi}_{(0,0)}^2}{12} + \frac{L^8}{r^4}\frac{\widetilde{\phi}_{(0,0)}^2}{48} \Bigg[\frac{1}{6}  \left(-9 \mu ^2 q^2 +(2+3 V_4) \widetilde{\phi}_{(0,0)}^2-36\frac{\phi_{(0,0)}}{\widetilde{\phi}_{(0,0)}} \right)\nonumber\\
 & & + \left((2+3 V_4) \widetilde{\phi}_{(0,0)}^2-3 \m^2q^2\right) \log\frac{r}{L}\Bigg] + \ldots \nonumber\\
f &\sim & 1 + \frac{L^8}{r^4}\left[f_{(4,0)} -\frac{1}{2} \left(\m q  \widetilde{\phi}_{(0,0)}\right)^2 \log \frac{r}{L}\right] + \ldots \nonumber\\
A_0 &\sim &\m + \frac{L^4}{r^2}\left(A_0{}_{(2,0)}-\frac{1}{4} q^2 \mu   \widetilde{\phi}_{(0,0)}^2 \log\frac{r}{L}\right) +\ldots \ .\label{eq:nbseries2}
\eea
The expansion in the $u=r_H^2/r^2$ coordinate will take the form
\bea
\phi &\sim & \alpha_- u^{1/2} + u^{3/2}\left(\alpha_+ \tilde{\alpha}_+ \log u\right)  +\ldots \nonumber\\
f &\sim  & 1 + u^2\left(f_4+\tilde{f}_4 \log u \right) + \ldots \nonumber\\
A_0 &\sim &  \m + \mu u\left( A_{0\, 2}+\tilde{A}_{0\,2}\log u\right) +\ldots \ . \label{eq:unbseries2}
\eea
Comparing the two expansions, one finds the following relations between the coefficients:
\bea
\alpha_- & = & \frac{L^2}{r_H}  \widetilde{\phi}_{(0,0)} \\
\tilde{\alpha}_+ & = & -\frac{1}{2}\frac{L^6}{r_H^3}\left[\widetilde{\phi}_{(0,0)}\left(\frac{1 }{2}\m^2 q^2 -\left(\frac{1}{3}+\frac{V_4}{2}\right) \widetilde{\phi}_{(0,0)}^2\right)\right] \\
\alpha_+ & = & \frac{L^6}{r_H^3}\left[ \phi_{(0,0)}-\widetilde{\phi}_{(0,0)}\left(\frac{1 }{2}\m^2 q^2 -\left(\frac{1}{3}+\frac{V_4}{2}\right) \widetilde{\phi}_{(0,0)}^2\right) \log \frac{r_H}{L} \right] \\
\tilde{f}_4 & = & \frac{1}{4}\frac{L^8}{r_H^4} \left(\m q  \widetilde{\phi}_{(0,0)}\right)^2 \\
 f_4 & = &  \frac{L^8}{r_H^4}\left[f_{(4,0)} +\frac{1}{2} \left(\m q  \widetilde{\phi}_{(0,0)}\right)^2 \log \frac{r_H}{L}\right] \\
 \tilde{A}_{0\,2} & = & \frac{1}{8}\frac{L^4}{r_H^2} q^2   \widetilde{\phi}_{(0,0)}^2 \\ 
 A_{0\, 2} & = & \frac{L^4}{r^2}\left(\frac{1}{\mu} A_0{}_{(2,0)}+\frac{1}{4} q^2    \widetilde{\phi}_{(0,0)}^2 \log\frac{r_H}{L}\right) \ .
\eea
Since the equations of motion in the $u$ coordinate are independent of the ratio $r_H/L$, we can extract the logarithmic dependence of the subleading terms explicitly,
\bea
\phi_{(0,0)} & = & \widehat{\phi}_{(0,0)}+\widetilde{\phi}_{(0,0)}\left(\frac{1 }{2}\m^2 q^2 -\left(\frac{1}{3}+\frac{V_4}{2}\right) \widetilde{\phi}_{(0,0)}^2\right) \log \frac{r_H}{L} \nonumber\\
f_{(4,0)} & = & \widehat{f}_{(4,0)}-\frac{1}{2} \left(\m q  \widetilde{\phi}_{(0,0)}\right)^2 \log \frac{r_H}{L} \nonumber\\
A_0{}_{(2,0)} & = & \widehat{A}_0{}_{(2,0)}-\frac{1}{4} q^2 \mu   \widetilde{\phi}_{(0,0)}^2 \log\frac{r_H}{L} \ , \label{eq:logdependence}
\eea
with
\begin{equation}\label{eq:mapru}
\widehat{\phi}_{(0,0)}=\alpha_+\frac{r_H^3}{L^6} , \ \ \widehat{f}_{(4,0)}=f_4\frac{r_H^4}{L^8} ,\ \ \widehat{A}_0{}_{(2,0)}=\mu A_{0\, 2}\frac{r_H^2}{L^4}  \ .
\end{equation}

\subsection{On-shell action}\label{app:onshell}

We can write Einstein's equations as
\begin{equation}
R_{MN} = T_{MN}^{(A)}+T_{MN}^\phi+\frac{1}{2} g_{MN}\left (\frac{L^2}{3}F^2+|D\phi|^2+\frac{5}{3}V\right) \ .
\end{equation}
From the trace of Einstein equations we find that the Ricci scalar is
\begin{equation}
R=\frac{L^2}{3}F^2+|D\phi|^2+\frac{5}{3}V \ .
\end{equation}
Therefore, the on-shell action \eqref{eq:bulkaction} is evaluated as
\begin{equation}
S_{\rm on-shell}=\frac{1}{16\pi G_5}\int d^5 x\sqrt{-g}\left[\frac{2}{3}V-\frac{2}{3}L^2 F^2 \right] \ .
\end{equation}
Let us now use that for our solutions
\begin{equation}
\Gamma^\alpha_{\mu\nu}=\Gamma^r_{r\nu}=\Gamma^\alpha_{rr}=0 \ , \ \Gamma^r_{\mu\nu}=-\frac{1}{\sqrt{g_{rr}}} K_{\mu\nu}\ , \ \Gamma^\alpha_{\mu r}=\sqrt{g_{rr}} K^\alpha_{\ \mu}\ , \ \Gamma^r_{rr}=\frac{1}{2}g^{rr}\partial_r g_{rr} \ ,
\end{equation}
where 
\begin{equation}
K_{\mu\nu}=\frac{1}{2\sqrt{g_{rr}}}\partial_r g_{\mu\nu} \ ,
\end{equation}
is the extrinsic curvature and $K^\alpha_{\ \mu}=g^{\alpha\beta}K_{\beta\mu}$, $K=g^{\mu\nu}K_{\mu\nu}$. We will also use that
\begin{equation}
\frac{\partial_r \sqrt{-g}}{\sqrt{-g}}=\Gamma^r_{rr}+\sqrt{g_{rr}}K\ .
\end{equation}
This allows us to write
\begin{equation}
g^{\mu\nu}R_{\mu\nu}=-\frac{1}{\sqrt{-g}}\partial_r\left( \frac{\sqrt{-g}}{\sqrt{g_{rr}}}K\right)=-\frac{1}{\sqrt{-g}}\partial_r\left( \sqrt{-\gamma} K\right) \ .
\end{equation}
We defined $\gamma_{\mu\nu}=g_{\mu\nu}$ as the boundary metric and used $\sqrt{-g}=\sqrt{g_{rr}}\sqrt{-\gamma}$.
On the other hand, from Einstein's equations,
\begin{equation}
g^{\mu\nu} R_{\mu\nu}=-\frac{2L^2}{3}F_{0 r} F^{0 r}+\frac{4}{3}V+q^2 g^{00}A_0^2\phi^2 \ ,
\end{equation}
where we only focused on the nonzero components on the solutions.

Solving for $V$ and introducing the result in the on-shell action, one gets
\begin{equation}
S_{\rm on-shell}=\frac{1}{16\pi G_5}\int d^5 x\sqrt{-g}\left[-\frac{1}{2\sqrt{-g}}\partial_r\left( \sqrt{-\gamma}K\right)-L^2 F_{r0}F^{r0}-\frac{q^2}{2} g^{00}A_0^2 \phi^2 \right]\ .
\end{equation}
Let us now use the equation of motion for the gauge field:
\begin{equation}
4L^2\partial_r\left( \sqrt{-g} F^{r0}\right)=2 q^2 \sqrt{-g} g^{00}A_0\phi^2 \ .
\end{equation}
We can then replace the $q^2$ term in the action by a derivative term and write the action as a total derivative:
\bea
S_{\rm on-shell} & = & \frac{1}{16\pi G_5}\int d^5 x\left[-\frac{1}{2}\partial_r\left(\sqrt{-\gamma}K\right)-L^2 \sqrt{-g} \partial_r A_0 F^{r0}-L^2A_0\partial_r\left( \sqrt{-g} F^{r0}\right)  \right]\nonumber\\
 & = & \frac{1}{16\pi G_5}\int d^5 x \,\partial_r \left[-\frac{1}{2}\sqrt{-\gamma}K-L^2 \sqrt{-g}  A_0 F^{r0} \right] \nonumber\\
 & = &-\frac{1}{16\pi G_5} \int d^4 x \left[ \frac{1}{2} \sqrt{-\gamma}K+L^2 \sqrt{-g}  A_0 F^{r0}\right]_{r=r_H}^{r=r_\Lambda} \ . \label{eq:onshellact}
\eea

\subsection{Connecting boundary and horizon values}

We can rewrite the equations for $f$ \eqref{eqf} and $A_0$  \eqref{eqA0} as
\bea
\left[ e^{4A} r^5 f'\right]'-L^4\left[4 e^{2A} r^3 (A_0')^2+2q^2 \frac{r e^{2A}A_0^2\phi^2 }{f}   \right] & = & 0 \\
\left[ e^{2A} r^3 A_0'\right]'-\frac{q^2}{2}  \frac{r  e^{2A} A_0\phi^2}{f} & = & 0 \ .
\eea
We can multiply the second equation by $-4L^4A_0$ and add it to the first one, yielding
\begin{equation}
\left[ e^{4A} r^5 f'-4L^4  e^{2A} r^3 A_0' A_0\right]'=0 \ .
\end{equation}
By introducing the function
\begin{equation}
\sigma(r)=\int_{r_H}^r dr_1 \frac{r_1  e^{2A}  A_0\phi^2}{f} \ , 
\end{equation}
the equations become 
\bea
\left[ e^{4A} r^5 f'-4L^4  e^{2A} r^3 A_0' A_0 \right]' & = & 0 \\
\left[ e^{2A} r^3 A_0'- \frac{q^2}{2}\sigma\right]' & = & 0 \ .
\eea
Let us further define
\begin{equation}\label{eq:bg}
\beta(r)=e^{4A} r^5 f'-4L^4  e^{2A} r^3 A_0' A_0, \ \  \gamma(r)= e^{2A} r^3 A_0'- \frac{q^2}{2}\sigma \ .
\end{equation}
By the equations above, these are independent of the radial coordinate, and it is convenient to evaluate them at the horizon $r\to r_H$,
\bea
\beta(r_H)=e^{4A(r_H)} r_H^5 f'(r_H) & \equiv & \beta_H \nonumber \\
 \gamma(r_H)=e^{2A(r_H)} r_H^3 A_0'(r_H) & \equiv & \gamma_H \ . \label{eq:bgH}
\eea

Note that the temperature and entropy density are given by
\begin{equation}
T=\frac{r_H^2 f'(r_H) }{4\pi L^2}e^{A(r_H)} \ , \ s=\frac{1}{4G_5} \frac{r_H^3}{L^3}e^{3 A(r_H)} \ .
\end{equation}
Then,
\begin{equation}\label{eq:betaH}
\beta_H=16\pi G_5  L^5 T s \ .
\end{equation}
The constant $\gamma_H$ is related to the charge density. We should expand now the solutions  \eqref{eq:bg} close to the boundary and compare with the values at the horizon \eqref{eq:bgH}. For $\Delta\neq 3$, $\sigma$ has the following expansion,
\begin{equation}\label{eq:solsbg}
	\sigma \simeq L^{16-4\Delta}\frac{\mu\widetilde{\phi}_{(0,0)}^2}{4(\Delta-3)}(r_H^{2\Delta-6}-r^{2\Delta-6})   +\sigma_b\ ,
\end{equation}
where $\sigma_b$ is a constant. The divergent term cancels out in the boundary expansion of $\gamma(r)$. Then, the matching of the constant terms at the boundary and the horizon leads to
\bea
A_0{}_{(2,0)} & = & -\frac{\gamma_H}{2L^4} +q^2\left[\frac{\mu   \widetilde{\phi}_{(0,0)}^2}{8(\Delta -3)}\left(\frac{r_H}{L^2}\right)^{2(\Delta-3)}  +\frac{\sigma_b}{2 L^4}\right] \nonumber\\
\qquad f_{(4,0)} & = & -\frac{1}{4}\frac{\beta_H}{L^8} +2\mu A_0{}_{(2,0)}\ . \label{eq:solsbge}
\eea
For $\Delta=3$, the expansion of $\sigma$ close to the boundary is
\begin{equation}\label{eq:solsbg2}
	\sigma\simeq \mu  L^4 \widetilde{\phi}_{(0,0)}^2 \left[\log \left(\frac{r}{L}\right)-\log\left(\frac{r_H}{L}\right)\right] + \sigma_b \ ,
\end{equation}
where again $\sigma_b$ is a constant. The divergent term cancels out in the boundary expansion of $\gamma(r)$, and the matching of the constant terms at the boundary and the horizon leads to
\bea 
A_0{}_{(2,0)} & = &  -\frac{\gamma_H}{2 L^4} - q^2 \left[ \frac{1}{8} \mu   \widetilde{\phi}_{(0,0)}^2 \left(1-2\log\frac{r_H}{L}\right)+\frac{\sigma_b}{4 L^4} \right] \\
f_{(4,0)} & = & -\frac{1}{4}\frac{\beta_H}{L^8} +2\mu A_0{}_{(2,0)}+\frac{1}{8}q^2 \mu^2 \widetilde{\phi}_{(0,0)}^2 \ . \label{eq:solsbge2}
\eea

\section{Holographic renormalization}\label{app:holoren}

In order to compute renormalized quantities we follow the holographic renormalization prescription  \cite{deHaro:2000vlm,Papadimitriou:2004rz,Papadimitriou:2005ii}. First, we introduce a cutoff in the radial direction $r_\Lambda$ such that the divergences of the bulk action \eqref{eq:bulkaction} are regulated. To this we should add the Gibbons-Hawking term in order to have a well-defined variational principle for the metric and boundary counterterms to remove the divergences. We distinguish between two cases, noninteger $\Delta$ and $\Delta=3$. We assume that the form of the metric is
\begin{equation}
ds^2=N^2 dr^2+\gamma_{\mu\nu}dx^\mu dx^\nu \ ,
\end{equation}
and define the extrinsic curvature and Brown-York tensors in the usual way:
\be 
\quad K_{\m\n} = \frac{1}{2N}\partial_r \gamma_{\m\n} \ , \ K = \gamma^{\m\n} K_{\m\n}\ , \ \Pi^{\m\n}_{BY} = \left( K^{\mu\nu}-\gamma^{\mu\nu}K\right) \ .
\ee

\subsection{Noninteger $\Delta$}

The boundary terms are
\bea
S_{GH} & = &\frac{1}{8\pi G_5}\int_{r=r_\Lambda} d^4 x\sqrt{-\gamma} K \nonumber\\ 
S_{c.t.} & = &\frac{1}{16\pi G_5}\int_{r=r_\Lambda} d^4 x\,\cL_{c.t.} \nonumber\\
\cL_{c.t.} & =&\sqrt{-\gamma}\left[ -\frac{6}{L}+\frac{\Delta-4}{L}\Phi^\dagger\Phi+\frac{L}{2(\Delta-3)} \gamma^{\mu\nu}(D_\mu\Phi)^\dagger D_\nu\Phi +\frac{1}{12L}\frac{(\Delta-4)^2}{\Delta-3}\left(\Phi^\dagger \Phi\right)^2 \right] \ .\nonumber
\eea 
The renormalized action is then
\begin{equation}
S_{ren}=\lim_{r_\Lambda\to\infty}\left[S_{bulk}+S_{GH}+S_{c.t.}\right]\ .
\end{equation}
The free energy is determined by the renormalized action $\beta V_3 \cF=-S_{ren}$ (where $\beta=1/T$ and $V_3$ is the spatial volume). Using the formula for the on-shell action \eqref{eq:onshellact} and the boundary expansions \eqref{eq:nbseries1} and the relation \eqref{eq:solsbge}, we find
\bea
\cF & = &\frac{1}{16\pi G_5}\left[ -\frac{1}{4}e^{4A(r_H)}\left(\frac{r_H}{L} \right)^5 f'(r_H)+2 L^3 \mu A_0{}_{(2,0)}+L^3(\Delta-4)(\Delta-2)  \widetilde{\phi}_{(0,0)} \phi_{(0,0)}\right] \nonumber\\
&= &-\frac{L^3}{16\pi G_5}\left[ -\frac{1}{4}\frac{\beta_H}{L^8}+2 \mu A_0{}_{(2,0)} +(\Delta-4)(\Delta-2)  \widetilde{\phi}_{(0,0)} \phi_{(0,0)}\right] \nonumber\\
&= &\frac{L^3}{16\pi G_5}\left[f_{(4,0)}+(\Delta-4)(\Delta-2)  \widetilde{\phi}_{(0,0)} \phi_{(0,0)}\right] \label{eq:freeenergy} \ .
\eea
Renormalized expectation values can be computed from variations of the action, using the same boundary terms,
\bea
\vev{T^{\m\n}} &=&  \frac{1}{8\pi G_5}\lim_{r \to \infty}\frac{r^2}{L^2}\left[- \sqrt{-\gamma}  \Pi_{BY}^{\m\n} +\frac{\delta \cL_{c.t.} }{\delta \gamma_{\mu\nu}} \right]  \label{eq:temren}\\
\vev{\mathcal{O}} &=& \frac{1}{16\pi G_5}\lim_{r\to \infty}\frac{L^{2(4-\Delta)}}{r^{4-\Delta}} \left[ -\sqrt{-g} g^{rr}\partial_r \phi +\frac{\delta \cL_{c.t.}}{\delta \Phi^\dagger}\right]\label{eq:oren} \\
\vev{J^\mu} &=& \frac{1}{16\pi G_5} \lim_{r\to \infty} \left[-4L^2\sqrt{-g} g^{rr} g^{\mu \alpha} F_{r\alpha}+\frac{\delta \cL_{c.t.}}{\delta A_\mu} \right]\ .\label{eq:jren}
\eea 
Introducing \Eq{eq:nbseries1} in the formulas \Eq{eq:temren}--\Eq{eq:jren}, we get 
\bea
\vev{T^{00}} &=& -\frac{L^3}{16 \pi G_5}\left[3 f_{(4,0)} - (\Delta-4) (\Delta -2) \widetilde{\phi}_{(0,0)} \phi_{(0,0)}\right] \label{eq:t00}\\
\vev{T^{ii}} &=& -\frac{L^3}{16 \pi G_5}\left[f_{(4,0)} + (\Delta-4) (\Delta -2) \widetilde{\phi}_{(0,0)} \phi_{(0,0)}\right] \label{eq:tii} \\
\vev{\mathcal{O}} &=& -\frac{L^3}{8 \pi G_5}(\Delta -2) \phi_{(0,0)} \label{eq:O}\\
\vev{J^0} &=& -\frac{L^3}{2\pi  G_5} A_0{}_{(2,0)} \ . \label{eq:j0}
\eea
Note that the Ward identity for the trace holds,
\be
\vev{T^\m\!_\m} = \vev{T^{\m\n}}\eta_{\m\n} = -(4-\Delta)\left(\vev{\mathcal{O}}\widetilde{\phi}_{(0,0)}^\dagger+\vev{\mathcal{O}^\dagger}\widetilde{\phi}_{(0,0)}\right)\ .
\ee

\subsection{$\Delta=3$}

The boundary terms are
\bea
S_{GH} & =  &\frac{1}{8\pi G_5}\int_{r=r_\Lambda} d^4 x\sqrt{-\gamma} K \nonumber\\ 
S_{c.t.} & =  &\frac{1}{16\pi G_5}\int_{r=r_\Lambda} d^4 x\,\cL_{c.t.}\nonumber\\
\cL_{c.t.} & = &\sqrt{-\gamma}\left[ -\frac{6}{L}-\frac{1}{L}\Phi^\dagger\Phi+L \left[\log\frac{r}{L}+\cW_1\right] \gamma^{\mu\nu}(D_\mu\Phi)^\dagger D_\nu\Phi \right. \nonumber\\
&&\left.+\frac{1}{L}\left[\left( \frac{1}{3}+\frac{V_4}{2}\right)\log\frac{r}{L}+\cW_2\right]\left(\Phi^\dagger \Phi\right)^2 \right] \ ,\nonumber
\eea
where $\cW_1$, $\cW_2$ are finite contributions. 
The renormalized action is then
\begin{equation}
S_{ren}=\lim_{r_\Lambda\to\infty}\left[S_{bulk}+S_{GH}+S_{c.t.}\right] \ .
\end{equation}
The free energy is determined by the renormalized action $\beta V_3 \cF=-S_{ren}$ (where $\beta=1/T$ and $V_3$ is the spatial volume). Using the formula for the on-shell action \eqref{eq:onshellact} and the boundary expansions \eqref{eq:nbseries2} and the relation \eqref{eq:solsbge2}, we find
\bea
\cF &= & \frac{1}{16\pi G_5}\left[ -\frac{1}{4}e^{4A(r_H)}\left(\frac{r_H}{L} \right)^5 f'(r_H)+2 L^3 \mu A_0{}_{(2,0)}-L^3  \widetilde{\phi}_{(0,0)} \phi_{(0,0)}\right.   \nonumber\\
& & \left.+L^3 \left(\frac{1}{4}+\cW_1 \right)q^2 \mu^2 \widetilde{\phi}_{(0,0)}^2-L^3\left(\frac{1}{12}+\frac{V_4}{8}+\cW_2 \right)\widetilde{\phi}_{(0,0)}^4 \right] \nonumber\\
&= &-\frac{L^3}{16\pi G_5}\left[ -\frac{1}{4}\frac{\beta_H}{L^8}+2 \mu A_0{}_{(2,0)} -  \widetilde{\phi}_{(0,0)} \phi_{(0,0)}+ \left(\frac{1}{4}+\cW_1 \right)q^2 \mu^2 \widetilde{\phi}_{(0,0)}^2-\left(\frac{1}{12}+\frac{V_4}{8}+\cW_2 \right)\widetilde{\phi}_{(0,0)}^4\right] \nonumber\\
&= &\frac{L^3}{16\pi G_5}\left[f_{(4,0)}-  \widetilde{\phi}_{(0,0)} \phi_{(0,0)}+ \left(\frac{1}{8}+\cW_1 \right)q^2 \mu^2 \widetilde{\phi}_{(0,0)}^2-\left(\frac{1}{12}+\frac{V_4}{8}+\cW_2 \right)\widetilde{\phi}_{(0,0)}^4\right] \label{eq:freeenergy2} \ .
\eea
Renormalized expectation values can be computed from variations of the action, using the same boundary  terms. Introducing \Eq{eq:nbseries2} in the formulas \Eq{eq:temren}--\Eq{eq:jren} (with $\Delta=3$), we get 
\bea
\notag\vev{T^{00}} &=& -\frac{L^3}{16 \pi G_5}\left[3 f_{(4,0)} +\widetilde{\phi}_{(0,0)} \phi_{(0,0)}+ \left(\frac{3}{8}+\cW_1 \right)q^2 \mu^2 \widetilde{\phi}_{(0,0)}^2\right.\\
 & &\left.+\left(\frac{1}{12}+\frac{V_4}{8}+\cW_2 \right)\widetilde{\phi}_{(0,0)}^4\right] \label{eq:t002}\\
\notag \vev{T^{ii}} &=& -\frac{L^3}{16 \pi G_5}\left[f_{(4,0)} - \widetilde{\phi}_{(0,0)} \phi_{(0,0)}+ \left(\frac{1}{8}+\cW_1 \right)q^2 \mu^2 \widetilde{\phi}_{(0,0)}^2\right.\\
& &\left.-\left(\frac{1}{12}+\frac{V_4}{8}+\cW_2 \right)\widetilde{\phi}_{(0,0)}^4\right]  \label{eq:tii2} \\
\vev{\mathcal{O}} &=& -\frac{L^3}{16 \pi G_5}\left[ 2\phi_{(0,0)}- \left(\frac{1}{2}+\cW_1 \right)q^2 \mu^2 \widetilde{\phi}_{(0,0)} +\left(\frac{1}{3}+\frac{V_4}{2}+2\cW_2 \right)\widetilde{\phi}_{(0,0)}^3\right] \label{eq:O2}\\
\vev{J^0} &=& -\frac{L^3}{16\pi  G_5}\left[8 A_0{}_{(2,0)}+ \left(1+2\cW_1 \right)q^2 \mu \widetilde{\phi}_{(0,0)}^2 \right]. \label{eq:j02}
\eea
The Ward identity for the trace has an anomalous contribution,
\be
\vev{T^\m\!_\m} = \vev{T^{\m\n}}\eta_{\m\n} = -\left(\vev{\mathcal{O}}\widetilde{\phi}_{(0,0)}^\dagger+\vev{\mathcal{O}^\dagger}\widetilde{\phi}_{(0,0)}\right)+\cA \ ,
\ee
where
\begin{equation}
\cA= -\frac{L^3}{16\pi  G_5}\left[\eta^{\mu\nu}(D_\mu  \widetilde{\phi}_{(0,0)})^\dagger D_\nu  \widetilde{\phi}_{(0,0)}+\left(\frac{1}{3}+\frac{V_4}{2} \right)\left( \widetilde{\phi}_{(0,0)}^\dagger  \widetilde{\phi}_{(0,0)}\right)^2\right] \ .
\end{equation}
It will be convenient to extract the explicit logarithmic dependence \eqref{eq:logdependence} and define the finite counterterms as
\begin{equation}
\cW_1=\log \left( \widetilde{\phi}_{(0,0)} L\right)+\kappa_1,\ \ \cW_2=\left(\frac{1}{3}+\frac{V_4}{2} \right)\log\left( \widetilde{\phi}_{(0,0)} L\right)+\kappa_2 \ .
\end{equation}
Then,
\bea
\notag\vev{T^{00}} &=& -\frac{L^3}{16 \pi G_5}\left[3 \widehat{f}_{(4,0)} -\left(\mu^2 q^2\widetilde{\phi}_{(0,0)}^2+\left(\frac{1}{3}+\frac{V_4}{2} \right)\widetilde{\phi}_{(0,0)}^4\right)\log\frac{r_H}{L^2\widetilde{\phi}_{(0,0)}}\right. \\
& &  \left.+\widetilde{\phi}_{(0,0)} \widehat{\phi}_{(0,0)}+ \left(\frac{3}{8}+\kappa_1 \right)q^2 \mu^2 \widetilde{\phi}_{(0,0)}^2+\left(\frac{1}{12}+\frac{V_4}{8}+\kappa_2 \right)\widetilde{\phi}_{(0,0)}^4\right] \label{eq:t003}\\
\notag \vev{T^{ii}} &=& -\frac{L^3}{16 \pi G_5}\left[\widehat{f}_{(4,0)} -\left(\mu^2 q^2\widetilde{\phi}_{(0,0)}^2-\left(\frac{1}{3}+\frac{V_4}{2} \right)\widetilde{\phi}_{(0,0)}^4\right)\log\frac{r_H}{L^2\widetilde{\phi}_{(0,0)}}\right.\\
& & \left. - \widetilde{\phi}_{(0,0)} \widehat{\phi}_{(0,0)}+ \left(\frac{1}{8}+\kappa_1 \right)q^2 \mu^2 \widetilde{\phi}_{(0,0)}^2-\left(\frac{1}{12}+\frac{V_4}{8}+\kappa_2 \right)\widetilde{\phi}_{(0,0)}^4\right]  \label{eq:tii3} \\
\notag \vev{\mathcal{O}} &=& -\frac{L^3}{16 \pi G_5}\left[ 2\widehat{\phi}_{(0,0)}+\left(\mu^2 q^2\widetilde{\phi}_{(0,0)}-2\left(\frac{1}{3}+\frac{V_4}{2} \right)\widetilde{\phi}_{(0,0)}^3\right)\log\frac{r_H}{L^2\widetilde{\phi}_{(0,0)}} \right.\\
& & \left.- \left(\frac{1}{2}+\kappa_1 \right)q^2 \mu^2 \widetilde{\phi}_{(0,0)} +\left(\frac{1}{3}+\frac{V_4}{2}+2\kappa_2 \right)\widetilde{\phi}_{(0,0)}^3\right], \label{eq:O3}\\
\vev{J^0} &=& -\frac{L^3}{16\pi  G_5}\left[8 \widehat{A}_0{}_{(2,0)}-2\mu q^2\widetilde{\phi}_{(0,0)}^2\log\frac{r_H}{L^2\widetilde{\phi}_{(0,0)}} + \left(1+2\kappa_1 \right)q^2 \mu \widetilde{\phi}_{(0,0)}^2 \right].\label{eq:j03}
\eea

\bibliographystyle{JHEP}

\bibliography{biblio2}

\end{document}